\documentclass[11pt]{article}

\usepackage{amsmath}
\usepackage{amsfonts}
\usepackage{latexsym}
\usepackage{epsfig}
\usepackage{color}
\if@twoside\oddsidemargin25mm
\else
\renewcommand{\d}{\mathrm{d}}

\DeclareMathSymbol{\mg}{\mathrel}{symbols}{"1D}

\newcommand{\ga}{\alpha}
\newcommand{\gb}{\beta}

\newcommand{\gd}{\delta}
\renewcommand{\ge}{\epsilon}

\newcommand{\gf}{\phi}

\newcommand{\gx}{\xi}
\newcommand{\gm}{\mu}
\newcommand{\gn}{\nu}

\newcommand{\gk}{\kappa}
\newcommand{\gl}{\lambda}
\newcommand{\gr}{\rho}

\newcommand{\gs}{\sigma}

\newcommand{\go}{\omega}
\newcommand{\gz}{\zeta}
\newcommand{\gp}{\pi}
\newcommand{\gps}{\psi}
\newcommand{\get}{\eta}
\newcommand{\gch}{\chi}
\newcommand{\gG}{\Gamma}
\newcommand{\gD}{\Delta}

\newcommand{\gL}{\Lambda}

\newcommand{\gTh}{\Theta}

\newcommand{\cA}{{\cal A}}

\newcommand{\cO}{{\cal O}}
\newcommand{\cP}{{\cal P}}

\newcommand{\cR}{{\cal R}}

\newcommand{\tT}{{\tilde T}}

\newcommand{\Tr}{\mbox{Tr}}
\newcommand{\tr}{\text{tr}}

\newcommand{\Id}{\text{\small 1}\hspace{-3.5pt}\text{1}}

\newcommand{\ra}{\rightarrow}

\newcommand{\der}{\partial}

\newcommand{\inv}{^{-1}}
\newcommand{\dsp}{\displaystyle}

\newcommand{\ubar}[1]{{\underline{#1}}}

\newcommand{\labl}[1]{\label{#1}}
\newcommand{\beq}{\begin{equation}}
\newcommand{\eeq}{\end{equation}}
\newcommand{\barr}{\begin{array}}
\newcommand{\earr}{\end{array}}
\newcommand{\equ}[1]{\begin{gather} #1 \end{gather}}
\newcommand{\equa}[1]{\begin{align} #1 \end{align}}

\newcommand{\tabu}[2]{\begin{tabular}{#1} #2 \end{tabular}}
\newcommand{\arry}[2]{\begin{array}{#1} #2 \end{array}}

\newcommand{\pmtrx}[1]{\begin{pmatrix} #1 \end{pmatrix}}

\newcommand{\non}{\nonumber}
\newcounter{oldcounter}

\newcommand{\bder}{\bar\partial}
\newcommand{\fZ}{\mathfrak{ Z}}
\newcommand{\bgch}{{\bar\chi}}
\newcommand{\tgps}{{\tilde \psi}}

\newcommand{\tgG}{{\tilde \Gamma}}

\newcommand{\Intr}{\mathbb{Z}}
\newcommand{\Cplx}{\mathbb{C}}
\newcommand{\Real}{\mathbb{R}}

\newcommand{\ba}[2]{\[\begin{array}{#2}\label{#1}}
\newcommand{\ea}{\end{array}\]}
\newcommand{\be}{\begin{equation}}
\newcommand{\ee}{\end{equation}}
\newcommand{\bea}{\begin{eqnarray}}
\newcommand{\eea}{\end{eqnarray}}

\newcommand{\E}[1]{\mathrm{E_{#1}}}
\newcommand{\U}[1]{\mathrm{U(#1)}}
\newcommand{\SU}[1]{\mathrm{SU(#1)}}
\newcommand{\SO}[1]{\mathrm{SO(#1)}}
\newcommand{\Spin}[1]{\mathrm{Spin(#1)}}

\newcommand{\USp}[1]{\mathrm{USp(#1)}}

\newcommand{\rep}[1]{\mathbf{#1}}
\newcommand{\crep}[1]{\overline{\rep{#1}}}

\begin{document}

\begin{flushright}
hep-th/0308076
\\
KUNS-1860 \\ 
UVIC-TH/03-08
\\
\end{flushright}
\vskip 2 cm
\begin{center}
{\Large {\bf 
Localization of heterotic anomalies on various hyper surfaces of 
$\boldsymbol{T^6/\Intr_4}$} 
}
\\[0pt]

\bigskip
\bigskip {\large
{\bf Stefan Groot Nibbelink$^{a,}$\footnote{
{{ {\ {\ {\ E-mail: grootnib@uvic.ca}}}}}}}, 
{\bf Mark Hillenbach$^{b,}$\footnote{
{{ {\ {\ {\ E-mail:   mark@th.physik.uni-bonn.de}}}}}}},
{\bf Tatsuo Kobayashi$^{c,}$\footnote{
{{ {\ {\ {\ E-mail: kobayash@gauge.scphys.kyoto-u.ac.jp}}}}}}} 
\\ 
{\bf and  Martin G.A.\ Walter$^{b,}$\footnote{
{{ {\ {\ {\ E-mail: walter@th.physik.uni-bonn.de}}}}}}}
\bigskip }\\[0pt]
\vspace{0.23cm}
${}^a$ {\it 
University of Victoria, Dept.\ of Physics \& Astronomy, \\
PO Box 3055 STN CSC, Victoria, BC, V8W 3P6 Canada.\footnote{\ \ \ 
 Address after August 22nd: {\it 
School of Physics \& Astronomy, 
University of Minnesota, 116 Church Street S.E.,
Minneapolis, MN 55455, USA.}
}\\
(CITA National Fellow)\\
} 
\vspace{0.23cm}
${}^b$ {\it  
Physikalisches Institut der Universit\"at Bonn, \\
Nussallee 12, 53115 Bonn, Germany.\\
}
\vspace{0.23cm}
${}^c${\it  
Department of Physics, Kyoto University, \\ 
Kyoto 606--8502, Japan. \\ 
}
\bigskip
\vspace{1.4cm} 
\end{center}
\subsection*{\centering Abstract}

We investigate the structure of local anomalies of heterotic 
$\E{8}\times \E{8}'$ theory on $T^6/\Intr_4$. We show that the
untwisted states lead to anomalies in ten, six and four dimensions. 
At each of the six dimensional fixed spaces of this orbifold the
twisted states ensure, that the anomalies factorize separately. As some
of these twisted states live on $T^2/\Intr_2$,
they give rise to four dimensional anomalies as well. At all
four dimensional fixed points at worst a single Abelian anomaly can
arise. Since the anomalies in all these dimensions factorize in a
universal way, they can be canceled simultaneously. In addition,
we show that for all $\U{1}$ factors at the four dimensional fixed
points at least logarithmically divergent Fayet--Ilopoulos tadpoles
are generated.

\newpage

\section{Introduction}
\labl{sc:intro}

In this paper we investigate the structure of local anomalies on the
orbifold $T^6/\Intr_4$ within the context of heterotic $\E{8}\times
\E{8}'$ string theory. Strings on orbifolds were discussed first by
the authors of refs.\ \cite{dixon_85,Dixon:1986jc} and with the
inclusion of non--trivial gauge field backgrounds, so--called Wilson
lines, in \cite{ibanez_87,Ibanez:1988pj,Font:1988tp}. More recently
there has also been a lot of attention to compactifications on
orbifolds in field theory. An important development was the
investigation of the shape of anomalies on 
orbifolds. First in ref.\ \cite{Arkani-Hamed:2001is} the anomalies on 
$S^1/\Intr_2$ were computed and it was found that they localize
at the fixed points of this orbifold. Afterwards, various groups
computed anomalies on the orbifolds $S^1/\Intr_2$, 
$S^1/\Intr_2 \times \Intr_2'$
\cite{Scrucca:2001eb,Barbieri:2002ic,Pilo:2002hu,GrootNibbelink:2002qp}.
More general anomaly investigations, that apply to higher dimensional
orbifolds, have been pursued in ref.\
\cite{Asaka:2002my,vonGersdorff:2003dt,GrootNibbelink:2003gd}.

The question of the shape of anomalies in the context of heterotic
string theory compactified on $T^6/\Intr_3$ has been investigated in ref.\ 
\cite{Gmeiner:2002es}. This six dimensional
orbifold has only zero dimensional fixed points, which may support
twisted states. It was found that non--Abelian anomalies never arise
at these four dimensional fixed points. However, at each fixed point a
single anomalous $\U{1}$ is possible, not necessarily the same at each
fixed point. In ref.\ \cite{GrootNibbelink:2003gb} it was shown, that
a local four dimensional remnant of the Green--Schwarz mechanism
\cite{Green:1984sg} cancels these Abelian anomalies.

The existence of a global anomalous $\U{1}$ is associated with the
generation of a Fayet--Iliopoulos tadpole, which leads to spontaneous
breaking of the global anomalous $\U{1}$  
\cite{Dine:1987xk,Atick:1987gy,Dine:1987gj}. 
The existence of Fayet--Iliopoulos tadpoles on orbifolds, like
$S^1/\Intr_2\times \Intr_2'$, was realized in
\cite{Ghilencea:2001bw} and the shapes of these tadpoles over such
orbifolds have been computed in refs.\  
\cite{Barbieri:2001cz,Scrucca:2001eb,Barbieri:2002ic}. 
In the heterotic models on $T^6/\Intr_3$ local Fayet--Iliopoulos
tadpoles are also generated  \cite{GrootNibbelink:2003gb}. 
However, still only a global one necessarily leads to spontaneous
breaking. The full consequences of the local structure of these
tadpoles have not been fully understood yet, but they may lead to
dynamical instabilities as was discussed in five dimensional models on
$S^1/\Intr_2$ \cite{GrootNibbelink:2002wv,GrootNibbelink:2002qp}.

This paper continues the investigation of the papers
\cite{Gmeiner:2002es,GrootNibbelink:2003gb} 
of local anomalies in the
context of the heterotic string. We have chosen to work on the
non--prime orbifold $T^6/\Intr_4$, because it is the simplest six
dimensional orbifold, which contains fixed hyper surfaces of various
dimensions. We confirm the expectation that on both four and six
dimensional fixed spaces anomalies localize. Furthermore, we show that
the Green--Schwarz mechanism can cancel the local four, six and ten
dimensional anomalies simultaneously. And in addition, we compute 
the Fayet--Iliopoulos tadpoles for $\U{1}$ factors at the four
dimensional fixed points.

The paper is organized as follows: In section \ref{sc:Z4geom} we
describe the geometry of the orbifold $T^6/\Intr_4$, focusing in
particular on the fixed point structure. Next, we
investigate the local spectra at these four and six dimensional fixed
hyper surfaces: This includes both, the projections of
untwisted states at the fixed points, as well as the twisted states
that may live there. To calculate the local structure of
anomalies, we first develop general orbifold traces in section
\ref{sc:traces}, and explain how they can be applied to
anomalies. After that, we collect the local anomaly contributions of both
untwisted and twisted states at the four and six dimensional hyper
surfaces. We describe the local version of the Green--Schwarz
mechanism which cancels these local factorized anomalies. In section
\ref{sc:FItadp} we calculate the tadpoles associated with the
(anomalous) $\U{1}$'s at the four dimensional fixed points. Our
conclusions have been collected in section \ref{sc:concl}. We have
attached three appendices to this work: Appendix \ref{sc:spinlight} is
devoted to a description of spinors in various relevant dimensions using
light--cone gauge. The next appendix gives some background on
supergravity multiplets in six dimensions. In appendix 
\ref{sc:WeylReflect} we describe how $\Intr_2$ and $\Intr_4$ gauge
shifts can be classified.

\section{Geometry of $\boldsymbol{T^6/\Intr_4}$}
\labl{sc:Z4geom}

\begin{figure}
\[
\raisebox{0ex}{\scalebox{0.8}{\mbox{\input{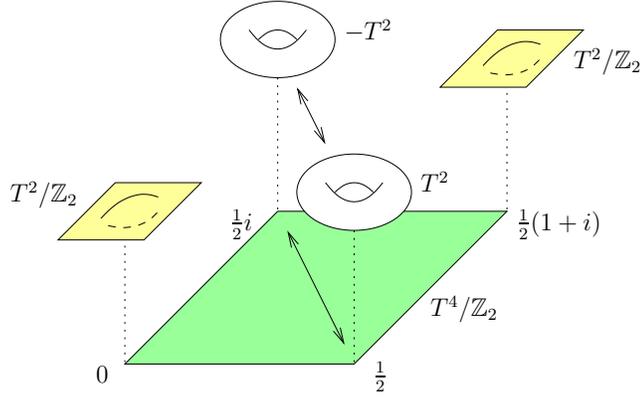}}}}
\]
\caption{
An impression of the two dimensional fixed hyper surfaces within the orbifold
$T^6/\Intr_4$ are displayed: the bottom square represents a part of
a two dimensional cross section (the $z_2$ plane) 
of the orbifold $T^4/\Intr_4$. Above
its fixed points $0$ and $\frac 12 (1+i)$ in the $z_1$ direction one
finds the orbifolds $T^2/\Intr_2$. Because of the identification of
the points $\frac 12$ and $\frac 12 i$ the two--tori above them are
mapped to each other via \eqref{identification6D}. 
} 
\labl{fig:hyper6D}
\end{figure}

We begin by reviewing the geometry of the orbifold $T^6/\Intr_4$,
a related discussion can be found in e.g.\ \cite{Ruan:2000uf}. 
Let $\gG$ be the lattice generated by 
$z_j \sim z_j + R_j$, $z_j \sim z_j + i\, R_j$ for $j =1,2,3$ on the
coordinates $z = (z_1, z_2, z_3)\in\Cplx^3$. This is the 
$\SO{5}\times\SO{5}\times\SU{2}\times\SU{2}$ lattice. 
(A general classification of orbifold compactification lattices can be
found in \cite{Kobayashi:1991mi,Kobayashi:1994rp}.)
We obtain the  torus 
$T^6 = \Cplx^3/\gG$ by dividing out this lattice.
The $\Intr_4$ twist operator $\gTh$ acts on the complex
coordinates as 
\equ{
\gTh (z_1, z_2, z_3) = (- z_1, i\, z_2, i\, z_3), 
\qquad 
\gTh^4 = \Id.
\labl{Z4twist}
}
For simplicity we have made the restriction to also only consider a square
torus in the first complex direction, even though the orbifolding does
not require this.

As the structure of fixed points is rather complicated, we introduce the
following notation: $\gz_{p\,q} = (p + i\, q)/2$ for 
$p, q \in \{0, 1 \}$. It is not hard to show that 
\equ{
i \,\gz_{p\, q} = \gz_{q\, p} - q, 
\quad 
i^2 \gz_{p\, q} = \gz_{p\, q} -p - i\, q, 
\quad 
i^3 \gz_{p\, q} = \gz_{q\, p} - i\, p.
\labl{FixedPointShifts}
}
The fixed points of the twists $\gTh$ and $\gTh^3$ are zero
dimensional. On these fixed points four dimensional states may arise
in the heterotic theory, as will be discussed in \ref{sc:4Dlocal}.  
There are 16 different $\Intr_4$ fixed points:
\equ{
\{ \fZ^4_{p\, q} \} = 
\{   
(R_1 \gz_{p_1q_1}, R_2 \gz_{p_2p_2}, R_3 \gz_{p_3p_3})
\}.
\labl{gThFixed}
}
Using the identities \eqref{FixedPointShifts} it is straightforward to
work out which lattice shifts are needed to make these fixed points
invariant within the covering space $\Cplx^3$ under the orbifold twist:
\equ{
\arry{rcl}{
\gTh\, \fZ^4_{p\, q} &=& \fZ^4_{p\, q} -
\bigl( (p_1 +i q_1) R_1, p_2 R_2, p_3 R_3 \bigr), 
\\[1ex]
\gTh^2 \fZ^4_{p\, q} &=& \fZ^4_{p\, q} -
\bigl(0, (1+i)p_2 R_2, (1+i)p_3 R_3 \bigr), 
\\[1ex]
\gTh^3 \fZ^4_{p\, q} &=& \fZ^4_{p\, q} -
\bigl( (p_1 +i q_1) R_1, i p_2 R_2, i p_3 R_3 \bigr).
}
\labl{gThFixedShifts}
}
These shifts are important since they distinguish the different fixed
points when Wilson lines are present: In section  \ref{sc:4Dlocal} we
will see that these shifts determine the local spectra at the fixed
points.

The  $\gTh^2$ fixed hyper surfaces take the form of $16$
disjoint $T^2$'s:
\equ{
\{(z_1, \fZ^2_{p\, q})~|~z_1 \in T^2 \} = 
\{   
(z_1, R_2 \gz_{p_2q_2}, R_3 \gz_{p_3q_3}) ~|~ z_1 \in T^2
\},
\labl{gTh2Fixed}
}
where the two--torus $T^2$ is defined by 
$z_1 \sim z_1 + R_1 \sim z_1 +i\,R_1$. Each two--torus 
$\{(z_1, \fZ^2_{p\, q})~|~z_1 \in T^2 \}$ can support six dimensional
twisted states, which will be determined in section \ref{sc:6Dlocal}. 
The local shifts that bring the fixed points back to themselves in the
covering space read 
\equ{
\gTh^2 \fZ^2_{p\, q} = \fZ^2_{p\, q} -
\bigl( (p_2 +i q_2) R_2, (p_3+i q_3) R_3 \bigr). 
\labl{gTh2FixedShifts}
}
Since on the fixed space of $\gTh^2$ the $\Intr_4$ twist acts
non--trivially, the embedding of this fixed space in the orbifold
$T^6/\Intr_4$ is somewhat more involved. (In figure \ref{fig:hyper6D}
we have given an artist's impression of the configuration.) 
The actions of $\gTh$ and $\gTh^3$ on this space take the form   
\equ{
\arry{rcl}{
\gTh\, (z_1, \fZ^2_{p\, q}) &=& 
(-z_1, \fZ^2_{q\, p}) - (0, q_2 R_2, q_3 R_3),
\\[1ex] 
\gTh^3 (z_1, \fZ^2_{p\, q}) &=& 
(-z_1, \fZ^2_{q\, p}) - i(0, p_2 R_2, p_3 R_3).
}
\labl{identification6D}
}
Notice that the order of $p$ and $q$ is interchanged, therefore it is
important to distinguish between the $\Intr_2$ fixed points with the
vectors $p$ and $q$ equal or not, denoted by $\fZ^2_{p=q}$ and 
$\fZ^2_{p\neq q}$, respectively. The twist $\gTh$ leaves $\fZ^2_{p=q}$
invariant, and hence the corresponding four two--tori  are
orbifolded to 
\equ{
\{   
(z_1, \fZ^2_{p=q}) ~|~ z_1 \in T^2/\Intr_2
\}  
=
\{   
(z_1, R_2 \gz_{p_2p_2}, R_3 \gz_{p_3p_3}) ~|~ z_1 \in T^2/\Intr_2
\}. 
\labl{gTh2eqFixed}
}
As each orbifold $T^2/\Intr_2$ itself has four fixed points 
$R_1 \gz_{p_1q_1}$, the fixed points of all four
disjunct orbifolds together is precisely the same as all fixed points
$\fZ^4_{p\,q}$ of the original orbifold $T^6/\Intr_4$.  
On the other 12 two--tori the twist $\gTh$ acts
freely; this leads to an identification of pairs of two--tori 
$(z_1, \fZ_{p\neq q}^2)$ and $(-z_1, \fZ_{q\neq p}^2)$ in the covering
space $\Cplx^3$ of the orbifold $T^6/\Intr_4$. In other words, within the
orbifold $T^6/\Intr_4$ these spaces really only consist of six two--tori. 
In the covering space $T^6$ these two--tori are indicated by 
\equ{
\{   
(z_1, \fZ^2_{p\neq q}) 
\oplus 
(-z_1, \fZ^2_{q\neq p}) 
~|~ z_1 \in T^2
\}. 
\labl{gTh2dfFixed}
}
As this is  a collection of two--tori, they do not have any orbifold
singularities. Notice that non of these two--tori contain the fixed points 
$\fZ^4_{p\, q}$.

We close our discussion with a few comments concerning the orbifold
$T^4/\Intr_2$. To gain insight in some properties of heterotic string
theory on $T^6/\Intr_4$, the relation to the four
dimensional orbifold $T^4/\Intr_2$ turns out to prove very useful. 
For this reason we collect here the essential geometrical properties
of this orbifold as well. We take $T^2 \times T^4$ to be described by
the same lattice as $T^6$ above. (For a general $T^4/\Intr_2$ orbifold
of course we do not need to take a square lattice in order that the
$\Intr_2$ can act consistently on it. For comparison
purposes between the theories on $T^6/\Intr_4$ and $T^4/\Intr_2$, we
restrict ourselves to square $T^4$'s only.) The $\Intr_2$
orbifold twist acts on the $T^4$ as  
$\gTh^2 (z_2, z_3) = ( -z_2, - z_3)$. It 
follows that the fixed points of $T^4/\Intr_2$ are given in
\eqref{gTh2Fixed}. The local shifts needed to bring these fixed points
back to themselves within the covering space $T^4$ are given in 
\eqref{gTh2FixedShifts}.

\section{Local ten, six and four dimensional string spectra}
\labl{sc:localspectra}

The central purpose of this section is to determine the string
spectrum of the heterotic $\E{8}\times \E{8}'$ theory on
$T^6/\Intr_4$. These models may contain arbitrary gauge shift and
Wilson lines. This makes the zero mode analysis of these models rather
complicated. However, as our analysis here will show the investigation
of the local spectra is relatively straightforward. 
In particular we are interested in the local twisted
and untwisted states at the six and four dimensional fixed hyper surfaces
discussed in section \ref{sc:Z4geom}.

\subsection{Ten dimensional states}
\labl{sc:10Dlocal}

The full ten dimensional spectrum of the heterotic $\E{8}\times \E{8}'$
string theory appears on the interior of the orbifold $T^6/\Intr_4$;
away from the fixed hyper surfaces. At the fixed hyper
surfaces a large number of states do not survive the local orbifold
projections, as we will discuss in later subsections. As this ten
dimensional spectrum is well--known, we will be brief at this point.

In table \ref{tb:10Dstates} we have summarized the 
ten dimensional zero mode spectrum of the
string on the interior of the orbifold $T^6/\Intr_4$ using 
light--cone gauge.  In this table $|0\rangle_{\widetilde{NS}}$ denotes
the right--moving Neveu--Schwarz vacuum, and 
\(
| S_1,\ldots S_4\rangle_{\widetilde{R}} 
\) 
the spinorial  Ramond vacuum with positive chirality: 
$S_i = \pm 1/2$ and $\prod S_i > 0$. Here it is used, that the
light--cone gauge automatically takes the ten dimensional Majorana
condition into account. (See appendix \ref{sc:spinlight} where we
review how ten (and lower) dimensional spinors can be represented on
the light--cone.) In table  \ref{tb:10Dstates}  $\ga^M_{-1}$ and
$\tgps^M_{-1/2}$ denote the creation operators of string world sheet
scalars and right--moving fermions, with $M,N$ spacetime indices.  
A light--cone spacetime vector, obtained from the $\widetilde{NS}$
vacuum, is indicated by 
\(
| \underline{\pm 1, 0,0,0} \rangle_{\widetilde{NS}} = 
\tgps^M_{-1/2} | 0 \rangle_{\widetilde{NS}}. 
\)
This notation will be extended to the twisted states in
sections \ref{sc:6Dlocal} and \ref{sc:4Dlocal}. And finally, 
$\rep{Ad}_{[0]}$ denotes the adjoint representation of
$\E{8}\times\E{8}'$, which consists of the Cartan 
generators $H_I$ of $\SO{16}\times \SO{16}'$, the adjoint roots 
$w = (\underline{\pm 1,\pm1, 0^6})$ and spinorial roots 
$w = (\pm 1/2,\ldots, \pm 1/2)$ (such that the product of the entries
is positive) of $\SO{16}$ and $\SO{16}'$. 
For more details we refer to \cite{pol_2,DiFrancesco:1997nk}. 
(The notation with the subscript $[0]$ will become useful, since we
will later define $\rep{Ad}_{[v]}$.)

\begin{table}
\begin{center}
\renewcommand{\arraystretch}{1.1}
\tabu{|c|}{
\hline 
ten dimensional heterotic spectrum 
\\ \hline\hline 
supergravity
\\ 
\tabu{p{7cm}cp{7cm}}{
\tabu{p{2.5cm}p{4.5cm}}{
$g_{MN}, B_{MN}, \gf ~:$ & 
$\ga^M_{-1} \tgps_{-\frac 12}^N |0\rangle_{\widetilde{NS}}$
}
& &
\tabu{p{1.5cm}p{5.5cm}}{
$\gps^M, \gl ~:$ & 
$ \ga^M_{-1} 
| \mbox{$ \frac{\pm1}2,  \frac{\pm1}2,  \frac{\pm1}2, \frac{\pm1}2$} \rangle_{\widetilde{R}~}$
}
}
\\ \hline 
super $\E{8}\times \E{8}'$ Yang--Mills
\\ 
\tabu{p{7cm}cp{7cm}}{
\tabu{p{2.5cm}p{4.5cm}}{
$A_{M} ~:$ 
&
$ | \rep{Ad}_{[0]} \rangle \otimes  
|  \underline{\pm 1, 0,0,0} \rangle_{\widetilde{NS}} $
}
& &
\tabu{p{1.5cm}p{5.5cm}}{
$\gch ~:$ 
&
$ | \rep{Ad}_{[0]} \rangle \otimes   
| \mbox{$ \frac{\pm1}2,  \frac{\pm1}2,  \frac{\pm1}2, \frac{\pm1}2$} \rangle_{\widetilde{R}~}$
}
}
\\ \hline 
}
\end{center}
\caption{
The ten dimensional zero mode spectrum of the heterotic string  is
identified using light--cone gauge ($M,N = 2,\dots 9$) in terms of string
oscillators and vacua.  
}
\labl{tb:10Dstates}
\end{table}

The $N=1$ supergravity part of the spectrum consists of a
graviton (i.e.\ the metric (perturbation)) $g_{MN}$, a
 scalar dilaton $\gf$, an anti--symmetric
tensor $B_{MN}$, and a left--handed gravitino $\psi^M$ 
and a right--handed dilatino $\gl$. These states are simply the
decomposition of the string states indicated in table
\ref{tb:10Dstates} in irreducible $\SO{8}$ representations. In
particular, the gravitino constraint 
$\gG_M \gps^M = 0$ implies that the chiralities of gravitino
and dilatino are opposite. This supergravity multiplet is
coupled to a ten dimensional super Yang--Mills theory, which contains 
an $\E{8}\times \E{8}'$ gauge field $A_M$ and a left--handed gaugino
$\gch$.

\subsection{Shift and Wilson lines} 
\labl{sc:Shift}

The transformation properties of the ten dimensional
supergravity sector are directly determined by specifying that the
orbifold $\gTh$ acts on an $\SO{8}$ representation as  
\equ{
\gTh \, | S_1,\ldots S_4 \rangle = e^{ 2 \pi i\, \gf_i S_i} \, 
| S_1,\ldots S_4 \rangle, 
\qquad 
\gf = \frac 14\, (0, \mbox{-}2, 1,1).  
}
Notice that this is consistent with the action on the spacetime
coordinates (also in light--cone gauge of course).

For the ten dimensional super Yang--Mills sector one can allow for
more choices: gauge shift $v$ and Wilson lines $a_j$ and
$\tilde a_j$ are free up to certain requirements which we recall below. 
Let $\hat\jmath$ denote the lattice vector of length
$R_j$ in the $j$th direction, we have for the $\E{8}\times\E{8}'$
gauge connection one--form: 
\equ{
A_1(z + \hat\jmath) = T_j \,A_1(z)\, T_j\inv, 
\quad 
A_1(z + i\, \hat\jmath) = \tT_j \,A_1(z)\, \tT_j\inv, 
\quad 
A_1(\gTh z) = U\, A_1(z) \, U\inv. 
}
The group elements $T_j$, $\tT_j$ and $U$ are assumed to be commuting;
they are all generated by the Cartan elements $H_I$ of $\SO{16}\times
\SO{16}' \subset \E{8} \times \E{8}'$. Their expressions
in terms of gauge shift and Wilson lines read:  
\equ{
T_j = e^{2\pi i\, a_j^I H_I},
\qquad 
\tT_j = e^{2\pi i\, \tilde a_j^I H_I}, 
\qquad 
U = e^{2\pi i\, v^I H_I}.
}
The other generators of the gauge group $\E{8}\times \E{8}'$ are
denoted by $E_w$, where $w$ are the roots of this algebra (described
above). In the Cartan--Weyl basis we have the canonical commutation 
and conjugation relations
\equ{
[H_I, E_w] = w_I H_I, 
\qquad 
e^{2\pi i\, t^I H_I}\,  E_w \, e^{-2\pi i\, t^I H_I} 
= e^{2\pi i\, t^I w_I} E_w,
}
for any $t^I \in \Real$. 
Because of the compatibility of the orbifold twist and the torus
periodicities \cite{Dixon:1986jc,Ibanez:1988pj,Kobayashi:1991mi,
Kobayashi:1994rp}, we find that: 
\equ{
\tilde a_j = a_j, 
\qquad 
4 v^I w_I = 2 a_j^I w_I = 2a_1^I w_I = 2 \tilde a_1^I w_I = 0 \mod 1
\labl{consWilson}
}
for $j=2,3$ and for all roots $w$. There are additional
conditions on gauge shift and Wilson lines coming from modular
invariance of the string theory. We discuss them below (in equation 
\eqref{modular}), since there we have introduced sufficient
ingredients to describe them naturally.

At the fixed hyper surfaces of the orbifold described in section
\ref{sc:Z4geom}, the combination of gauge shift and Wilson lines can
lead to different local projections of the untwisted gauge
states, i.e. gauge fields and gauginos \cite{Gmeiner:2002es}. 
At the fixed points $\fZ^4_{p\, q}$ and 
$\fZ^2_{p\, q}$ the gauge field one--form satisfies  
\equ{
A(\fZ^4_{p\, q}) = R^4_{p\, q} A(\fZ^4_{p\, q}) (R^4_{p\, q})\inv, 
\qquad 
A(z_1, \fZ^2_{p\, q}) = R^2_{p\, q} A(z_1,\fZ^2_{p\, q}) (R^2_{p\,q})\inv. 
}
The second condition applies to the pairwise identified two--tori and
the interior of the orbifolded two--tori. Since the fixed points of
these orbifolded two--tori are the fixed points $\fZ^4_{p\, q}$, there
the first requirement is again obtained. The local projection matrices
$R^4_{p\, q} = \exp({2 \pi i\, v^{4\ I}_{p\, q} H_I})$ 
and $R^2_{p\, q}= \exp({2 \pi i\, v^{2\ I}_{p\, q} H_I})$
can be expressed in terms of the local shift vectors 
\equ{
v^{4}_{p\, q} = p_1 a_1 + q_1 \tilde a_1 + p_2 a_2 + p_3 a_3 + v, 
\quad 
v^{2}_{p\, q} = (p_2+q_2) a_2 + (p_3 + q_3) a_3 + 2 v. 
\labl{LocalShifts}
}
These local shifts also determine the four and six dimensional twisted
states as we describe below. At the two tori which are identified, the
projections should of course be the same: This is indeed the case,
since $v^{2}_{p\, q} = v^{2}_{q\, p}$. Moreover, notice that because 
$2a_2^I w_I  = 2 a_3^I w_I = 0  \mod 1$, all local shifts 
at the four orbifolds $T^2/\Intr_2$ are equal: $v^2_{p=q} = 2v$ (up to
lattice shifts).

Because at all fixed (four and six dimensional) hyper surfaces 
the theory should correspond to a consistent string model,  all these
local shift vectors need to satisfy the modular invariance conditions 
\equ{
\Intr_4~: ~~ 4 \Bigl( \gf^2 - ( v^{4}_{p\, q})^2 \Bigr) = 0 \mod 2, 
\quad 
\Intr_2~:~~ 2 \Bigl( (2\gf)^2 - ( v^{2}_{p\, q})^2 \Bigr) = 0 \mod 2.
\labl{modular}
}
Not all these conditions are independent: The $\Intr_2$ level matching
conditions  for $v^2_{p=q}$ are automatically satisfied, provided that  
all $v^4_{p\,q}$ fulfill the $\Intr_4$ conditions. However, the
$\Intr_2$ conditions  for $v^2_{p\neq q}$ give extra independent
relations in general. Using that the conditions in \eqref{modular}
hold for all $p_i, q_i = 0,1$ with $i = 1,2,3$ we find the
requirements 
\equ{
\arry{c}{
2 \bigl( \gf^2 - v^2 \bigr) = a_2^2 = a_3^2 = 2\, a_1^2 = 2\, \tilde a_1^2 
= 0 \mod 1, 
\\[1ex] 
2\, a_2 a_3 = 4\, a_1 a_2 = 4\, a_1 a_3 
= 4\, \tilde a_1 a_2 = 4\, \tilde a_1 a_3  = 0 \mod 1, 
\\[1ex]
4\, v \tilde a_1 = 4\, v a_1 = 4\, v a_2 = 4\, v a_3 = 0 \mod 1.  
}
}

For the related orbifold models on $T^4/\Intr_2$ we take spacetime twist
and gauge shifts to be $2 \gf$ and $2 v$, respectively. Furthermore
the Wilson lines in the real and imaginary $z_2$ and $z_3$ directions,
are respectively $a_2$ and $a_3$. Hence, we find the same local gauge
shift $v^2_{p\, q}$ as given in \eqref{LocalShifts} with the modular
invariance requirement \eqref{modular}.

\subsection{Local six dimensional spectra}
\labl{sc:6Dlocal}

\begin{table}
\begin{center}
\renewcommand{\arraystretch}{1.1}
\tabu{|c|}{
\hline 
local six dimensional spectrum 
\\ \hline\hline 
untwisted (orbifolded 10 D) states
\\ \hline\hline 
supergravity + tensor 
\\ 
\tabu{p{7cm}c p{7cm}}{
\tabu{p{2.5cm}p{4.5cm}}{
$g_{mn}, B_{mn}, \gf ~:$ & 
$\ga^m_{-1} \tgps_{-\frac 12}^n |0\rangle_{\widetilde{NS}}$
}
& &
\tabu{p{1.5cm}p{5.5cm}}{
$\gps^m, \gl ~:$ & 
$ \ga^m_{-1} | \mbox{$\frac {\ga}2$}, \mbox{$\frac {\ga}2$}, 
\mbox{$\frac {\gb}2$},\mbox{$\frac {\gb}2$} \rangle_{\widetilde{R}~}$
}
}
\\[-2ex] 
neutral hyper 
\\ 
\tabu{p{7cm}c p{7cm}}{
\tabu{p{2.5cm}p{4.5cm}}{
$q_{ab} ~:$ & 
$\ga^a_{-1} \tgps_{-\frac 12}^b |0\rangle_{\widetilde{NS}}$
}
& &
\tabu{p{1.5cm}p{5.5cm}}{
$\gz_{ab} ~:$ & 
$ \ga^a_{-1} | \mbox{$\frac {\ga}2$}, \mbox{-$\frac {\ga}2$}, 
\mbox{$\frac {\gb}2$},\mbox{-$\frac {\gb}2$} \rangle_{\widetilde{R}~}$
}
}
\\ \hline 
vector 
\\ 
\tabu{p{7cm}cp{7cm}}{
\tabu{p{2.5cm}p{4.5cm}}{
$A_{m} ~:$ 
&
$ | \rep{Ad}_{[V_2]} \rangle \otimes 
| \underline{\pm 1, 0},0,0 \rangle_{\widetilde{NS}} $
}
& &
\tabu{p{1.5cm}p{5.5cm}}{
$\gch ~:$ 
&
$ | \rep{Ad}_{[V_2]} \rangle \otimes   
 | \mbox{$\frac {\ga}2$}, \mbox{$\frac {\ga}2$}, 
\mbox{$\frac {\gb}2$},\mbox{$\frac {\gb}2$} \rangle_{\widetilde{R}~}$
}
}
\\[-2ex] 
charged hyper  
\\ 
\tabu{p{7cm}cp{7cm}}{
\tabu{p{2.5cm}p{4.5cm}}{
$A^{\rep{R}_{[V_2]}} ~:$ 
&
$ | \rep{R}_{[V_2]} \rangle \otimes 
| 0,0,\underline{\pm 1,0} \rangle_{\widetilde{NS}} $
}
& &
\tabu{p{1.5cm}p{5.5cm}}{
$\gz^{\rep{R}_{[V_2]}} ~:$ 
&
$ | \rep{R}_{[V_2]} \rangle \otimes   
 | \mbox{$\frac {\ga}2$}, \mbox{-$\frac {\ga}2$}, 
\mbox{$\frac {\gb}2$},\mbox{-$\frac {\gb}2$} \rangle_{\widetilde{R}~}$
}
}
\\ \hline \hline 
twisted states 
\\ \hline \hline
charged $\SU{2}_H$ singlet hyper 
\\ 
\tabu{p{7cm}cp{7cm}}{
\tabu{p{2.5cm}p{4.5cm}}{
$B^{\rep{S}_{[V_2]}} ~:$ 
&
$ | \rep{S}_{[V_2]} \rangle \otimes 
| 0,0, \mbox{$\frac {\ga}2$}, \mbox{$\frac {\ga}2$} \rangle_{\widetilde{NS}} $
}
& &
\tabu{p{1.5cm}p{5.5cm}}{
$\gx^{\rep{S}_{[V_2]}} ~:$ 
&
$ | \rep{S}_{[V_2]} \rangle \otimes   
 | \mbox{$\frac {\ga}2$}, \mbox{-$\frac {\ga}2$},0,0 \rangle_{\widetilde{R}~}$
}
}
\\[-2ex] 
charged $\SU{2}_H$ doublet hyper 
\\ 
\tabu{p{7cm}cp{7cm}}{
\tabu{p{2.5cm}p{4.5cm}}{
$B^{\rep{D}_{[V_2]}} ~:$ 
&
$ | \rep{D}_{[V_2]} \rangle \otimes \ga^a_{-\frac 12} 
| 0,0, \mbox{$\frac {\ga}2$}, \mbox{$\frac {\ga}2$} \rangle_{\widetilde{NS}} $
}
& &
\tabu{p{1.5cm}p{5.5cm}}{
$\gx^{\rep{D}_{[V_2]}} ~:$ 
&
$ | \rep{D}_{[V_2]} \rangle \otimes  \ga^a_{-\frac 12} 
 | \mbox{$\frac {\ga}2$}, \mbox{-$\frac {\ga}2$},0,0 \rangle_{\widetilde{R}~}$
}
}
\\ \hline 
}
\end{center}
\caption{
The local six dimensional spectrum of the heterotic string 
is given in terms of string oscillators and vacua. This spectrum is
situated at the fixed points of $T^4/\Intr_2$ and six dimensional
fixed hyper surfaces of $T^6/\Intr_4$. The
following indices are used: light--cone: $m,n = 2,\dots 5$,
internal $T^4$: $a,b = i, \bar i = 2, 3, \ubar 2, \ubar 3$, 
and spinor: $\ga,\gb=\pm$. (To make the multiplet structure more
manifest we have used the identification of internal space and spinor
indices.)
}
\labl{tb:6Dspec}
\end{table}

\begin{table}
\begin{center}
\renewcommand{\arraystretch}{1.1}
\begin{tabular}{|c|c|c|c|} \hline
no. & shift & gauge group & {untwisted matter}
\\
$ 2(V_2)^2 $ & $2 V_{2}$ & $\rep{Ad}_{[V_2]}$ & $\rep{R}_{[V_2]}$  
\\ \hline \hline 
$0$ & $(00000000)$ & $\E{8}$  & nothing 
\\ \hline
$1, 3$  & $(11000000)$ & $\E{7} \times \SU{2}$  
& $(\rep{56},\rep{2})$ 
\\ \hline
$2$ & $(20000000)$ & $\SO{16}$ &  $\rep{128}_s$      
\\ \hline 
\end{tabular}
\end{center}
\caption{The resulting gauge groups and six dimensional untwisted
matter representation are given for representatives of the possible
$\Intr_2$ gauge shifts. General shifts are classified by computing
$2V_2^2 \mod 4$ as is discussed in appendix \ref{sc:Z2class}.  
(Since gauge shifts with $2V_2^2 = 1,3 \mod 4$ are just
related to each other via some lattice shift, we will use only the
symbol ``$1$'' for classification purposes.) 
}  
\labl{tb:class-2-U}
\end{table}

The discussion of the local shift vectors in the preceding section,
allows us to make an inventory of the local states at the fixed hyper
surfaces of the orbifold $T^6/\Intr_4$. In this subsection we consider
the six dimensional states. These states live on four dimensional
Minkowski space times, either two identified two--tori, or the orbifold
$T^2/\Intr_2$. However, as far as the classification of six
dimensional states is concerned, the local spectra at the fixed points
are completely determined by the local $\Intr_2$ shift vectors $v^{2}_{p\, q}$.
Therefore, for a giving fixed six dimensional hyper surface
$\fZ^2_{p\,q}$ of $T^6/\Intr_4$,  we can consider an equivalent pure
orbifold model (i.e. without Wilson lines) on $T^4/\Intr_2$ with this
gauge shift to determine the 
local spectrum there. The method of using equivalent models to
determine the local spectra with Wilson lines present, was
employed for the $T^6/\Intr_3$ orbifold in \cite{Gmeiner:2002es}. 
However, we should emphasize here, that in the 
present case we make identifications between spectra at fixed points
of two theories on different orbifolds. Although for spectra this
method works, one should be aware of important differences when
computing anomalies (and other traces of local operators) on these
different spaces. We will return to this important issue in section
\ref{sc:traces}, where traces on these orbifolds are evaluated.

Let $V_2$ be the gauge shift of a pure orbifold $T^4/\Intr_2$
model with spacetime shift $2 \gf$. 
(We will denote the classifying shifts with capital letters:
$V_2$ and $V_4$ for the $\Intr_2$ and $\Intr_4$ orbifolding,
respectively.) 
The string spectrum can be divided in untwisted and twisted
states. The untwisted states can be understood as those ten
dimensional states that survive the $\Intr_2$ orbifolding, or can be
obtained by a direct string calculation of untwisted six dimensional
zero modes. In any case, these states are invariant
under the orbifold action. Using the $\SO{8}$ light--cone
representations given  in section \ref{sc:10Dlocal} and their six
dimensional decompositions  \eqref{SO8branching} of appendix
\ref{sc:6Dmulti}, we can classify the untwisted states according to
the six dimensional supergravity multiplets reviewed in table
\ref{tb:6DFermions} of the same appendix. For the states coming from
the ten dimensional supergravity sector, this is rather
straightforward. The ten dimensional gauge multiplet states 
give rise to six dimensional gauge multiplets in the adjoint
$\rep{Ad}_{[V_2]}$ of the gauge group $G_{[V_2]}$ unbroken by the 
orbifolding, and to charged hyper multiplets in representation
$\rep{R}_{[V_2]}$. These representations are defined by  
\equ{
\arry{l}{
\rep{Ad}_{[V_2]} = \{ H_I \} \oplus 
\{ \E{8}\ \text{roots}\ w ~|~ V_2^I \, w_I = 0 \mod 1 \}, 
\\[2ex]
\rep{R}_{[V_2]} \ \ = \{\E{8}\ \text{roots}\ w ~|~ V_2^I \, w_I 
= \mbox{$\frac 12$} \mod 1 \}.
}
\labl{defAdR}
}
These definitions only apply to a single $\E{8}$, but can, of
course, be easily extended to $\E{8}\times\E{8}'$. (In the following we
use this notation for both situations, assuming that the context makes
clear whether one is concerned with a single or both $\E{8}$'s.) In table
\ref{tb:6Dspec} we have summarized the full untwisted spectrum of the
theory in six dimensions and indicated to which string states they
correspond. Since the untwisted matter states contain $4+4$ degrees of
freedom, they naturally fall into hyper multiplets in gauge group
representation $\rep{R}_{[V_2]}$. To emphasize, this spectrum
corresponds to the local spectrum at the fixed points of $T^4/\Intr_2$
but likewise to the spectrum at the $T^2$'s and $T^2/\Intr_2$'s within
$T^6/\Intr_4$.

In addition to the untwisted spectrum, there exist twisted modes, 
which correspond to additional string states that are massless
because of the orbifolding. As we will need similar formulae for the
$\Intr_4$ twisted states, we describe here  the masslessness
conditions for a $k$th twisted sector of a $\Intr_N$ orbifold. (The $0$th
twisted sector gives the untwisted spectrum which we already
characterized.) The masslessness conditions for the $k$th twisted sector of
a $\Intr_N$ orbifold with gauge shift $V$ and spacetime twist $\gf$
read 
\equ{
\mbox{$\frac 12$} (w-k\,  V)^2+N^{(k)}_L + c_{(k)}-1=0,
\qquad 
\mbox{$\frac 12$} 
(\omega + k \,\gf)^2 + N^{(k)}_R+ c_{(k)} - \frac 12=0,
\label{Tw-Massless}
}
with 
\equ{ 
c_k \equiv \mbox{$\frac 12$} \sum_{i=1}^4\eta^i_{(k)}(1-\eta^i_{(k)}), 
\qquad 
 \eta^i_{(k)} \equiv |k\phi^i|-{\rm Int}|k\phi^i|. 
}
The left and right--moving oscillator numbers $N^{(k)}_L$
and $N^{(k)}_R$ are fractionally quantized in terms of the modding of
the various left-- and right--moving world sheet fields. There are
only a few possibilities for these oscillator numbers, since for
massless states they are bounded from above by $1 - c_{(k)}$ 
and $1/2 - c_{(k)}$, respectively. 
The weights $w$ in the ${\E{8}\times\E{8}'}$ root
lattice, which satisfy the above requirements, define left--moving
vacua $| w \rangle$. Likewise, the 
weights $\go$ of the $\SO{8}$ root lattice (possibly shifted by the
spinorial root $(\frac 12^4)$) determine the right--moving vacua
$| \go \rangle$. In addition the tensor
products of the left-- and right--moving states need to fulfill the
generalized GSO projection:   
\equ{
\cP^{(k)} \, 
\exp 2 \pi i\, \Bigr\{\frac {k}2 \Bigl( V^2 - \phi^2 \Bigr) 
+ (w -k\, V) V + (\omega + k\, \phi) \phi \Bigr\} = 1,
\labl{GSOvac}
}
where the $\cP^{(k)}$ denote the phases due to (fractional) oscillator
contributions.

Back to the six dimensional twisted states of $\Intr_2$. Since 
$c_{(1)} = 1/4$, the only possibility in the right--moving sector is
$\smash{N^{(1)}_R = 0}$ so that $( \go + 2\gf)^2 = 1/2$. 
The GSO projection then only allows the vacuum states 
$| 0,0, \mbox{$\frac {\ga}2$}, \mbox{$\frac {\ga}2$} \rangle_{\widetilde{NS}}$
and 
$| \mbox{$\frac {\ga}2$}, \mbox{-$\frac {\ga}2$},0,0 \rangle_{\widetilde{R}~}$.
For the left--movers we have two options:
\equ{
N^{(1)}_L = 0: ~~ 
\rep{S}_{[V_2]} = \{ w ~|~  
\mbox{$\frac 12$} (w - V_2)^2 = \mbox{$\frac 34$} \}; 
\qquad 
N^{(1)}_L = \mbox{$\frac 12$}: 
~~ 
\rep{D}_{[V_2]} = \{w ~|~  
\mbox{$\frac 12$}(w - V_2)^2 = \mbox{$\frac 14$} \}. 
\labl{RepCond}
}
Since the latter states are obtained by acting with the creation
operators $\ga^a_{-1/2}$ on the vacuum, these states transform as 
\(
(\rep{2}, \rep{2}) \otimes (\rep{2},\rep{1}) = 
(\rep{3},\rep{2}) \oplus (\rep{1},\rep{2})
\)
under $\SU{2}_R \times \SU{2}_H$. In table \ref{tb:6Dspec} we have
summarized this twisted spectrum as well.

\begin{table}
\begin{center}
\renewcommand{\arraystretch}{1.1}
\begin{tabular}{|c|l|c c |}\hline
shift  & \multicolumn{1}{|c|}{gauge group}  &  
\multicolumn{2}{|c|}{local twisted matter}
\\ 
$V_{2}$ &  \multicolumn{1}{|c|}{$G_{[V_2]}$} & 
$\rep{S}_{[V_2]}$ & $\rep{D}_{[V_2]}$ 
\\ \hline \hline 
 $(1;0)$ & $\E{7}\times \SU{2} ~\times ~\E{8}'$    
& 
$(\rep{56},\rep{1})(\rep{1})' $ & $4 (\rep{1},\rep{2})(\rep{1})' $
\\ \hline
$(1;2)$ & $\E{7} \times \SU{2}~\times~ \SO{16}'$ & 
$(\rep{1},\rep{2})(\rep{16}_v)'$ & 
\\ \hline
\end{tabular}
\end{center}
\caption{There are two modular invariant combinations of $\Intr_2$
gauge shifts. The labels $(V_2; V_2')$ correspond to the first column
of table \ref{tb:class-2-U}.  The resulting gauge group and the
twisted matter at a single fixed point is given.}
\labl{tb:class-2-T}
\end{table}

Let us now discuss what the possible local six dimensional models
are, by specifying modular invariant gauge shifts. 
In appendix \ref{sc:Z2class} we show that there are essentially three
different $\Intr_2$ gauge shift vectors within a single $\E{8}$, which can
be classified by the value of $2(V_2)^2$, and are listed in table
\ref{tb:class-2-U}. We label these possible gauge shifts by their value
for $2(V_2)^2$. This could in principle lead to six different models,
when the gauge shifts for both $\E{8}$'s are combined. However, the
$\Intr_2$ level matching condition \eqref{modular} only allows
essentially two combinations, which are given in  
table \ref{tb:class-2-T}. As the arising gauge group already determines the
untwisted states according to table \ref{tb:class-2-U}, we
have only given the gauge representations of the twisted states in
table \ref{tb:class-2-T}. The twisted states in representation 
$\rep{S}_{[V_2]}$ correspond only to $2+2$ degrees of freedom
according to table \ref{tb:6Dspec}, which would be too little to fill
six dimensional hyper multiplets. But from table \ref{tb:class-2-T} we
may read off that all representations $\rep{S}_{[V_2]}$ are pseudo real,
hence as discussed in appendix \ref{sc:6Dmulti} these representation
come with an anti--symmetric matrix, which allows one to form 
hyper multiplets with half their canonical degrees of freedom. For the
other twisted states in representation $\rep{D}_{[V_2]}$ there are two
ways to form such half--hyper multiplets, since they fall in
a doublet of the $\SU{2}_H$ and the $\SU{2}$ gauge group, see 
tables \ref{tb:6Dspec} and \ref{tb:class-2-T}.

\subsection{Local four dimensional spectra}
\labl{sc:4Dlocal}

\begin{table}
\begin{center}
\renewcommand{\arraystretch}{1.1}
\tabu{|c|}{
\hline 
local four dimensional spectrum 
\\ \hline\hline 
untwisted (orbifolded 10 D) states
\\ \hline\hline 
vector 
\\ 
\tabu{p{7cm}cp{7cm}}{
\tabu{p{2.5cm}p{4.5cm}}{
$A_{\gm} ~:$ 
&
$ | \rep{Ad}_{[V_4]} \rangle \otimes 
| \pm 1,0,0,0 \rangle_{\widetilde{NS}} $
}
& &
\tabu{p{1.5cm}p{5.5cm}}{
$\gch ~:$ 
&
$ | \rep{Ad}_{[V_4]} \rangle \otimes   
 | \mbox{$\frac {\ga}2$}, \mbox{$\frac {\ga}2$}, 
\mbox{$\frac {\ga}2$},\mbox{$\frac {\ga}2$} \rangle_{\widetilde{R}~}$
}
}
\\[-2ex] 
chiral 
\\ 
\tabu{p{7cm}cp{7cm}}{
\tabu{p{2.5cm}p{4.5cm}}{
$A_1^{\rep{R}_{[V_4]}} ~:$ 
&
$ | \rep{R}_{[V_4]} \rangle \otimes 
| 0,{\pm 1},0,0 \rangle_{\widetilde{NS}} $
}
& &
\tabu{p{1.5cm}p{5.5cm}}{
$\gz_1^{\rep{R}_{[V_4]}} ~:$ 
&
$ | \rep{R}_{[V_2]} \rangle \otimes   
 | \mbox{$\frac {\ga}2$}, \mbox{$\frac {\ga}2$}, 
\mbox{-$\frac {\ga}2$},\mbox{-$\frac {\ga}2$} \rangle_{\widetilde{R}~}$
}
}
\\[-2ex] 
chiral 
\\ 
\tabu{p{7cm}cp{7cm}}{
\tabu{p{2.5cm}p{4.5cm}}{
$A_j^{\rep{r}_{[V_4]}} ~:$ 
&
$ | \rep{r}_{[V_4]} \rangle \otimes 
| 0,0,\underline{\pm 1,0} \rangle_{\widetilde{NS}} $
}
& &
\tabu{p{1.5cm}p{5.5cm}}{
$\gz_j^{\rep{r}_{[V_4]}} ~:$ 
&
$ | \rep{r}_{[V_2]} \rangle \otimes   
 | \mbox{$\frac {1}2$}, \mbox{-$\frac {1}2$}, 
\mbox{$\frac {\ga}2$},\mbox{-$\frac {\ga}2$} \rangle_{\widetilde{R}~}$
}
}
\\ \hline \hline 
double twisted (orbifolded 6D) states 
\\ \hline \hline
chiral 
\\ 
\tabu{p{7cm}cp{7cm}}{
\tabu{p{2.5cm}p{4.5cm}}{
$B^{\rep{T_2}_{[V_2]}} ~:$ 
&
$ | \rep{T_2}_{[V_4]} \rangle \otimes 
| 0,0, \mbox{$\frac {1}2$}, \mbox{$\frac {1}2$} \rangle_{\widetilde{NS}} $
}
& &
\tabu{p{1.5cm}p{5.5cm}}{
$\gx^{\rep{T_2}_{[V_2]}} ~:$ 
&
$ | \rep{T_2}_{[V_2]} \rangle \otimes   
 | \mbox{$\frac {1}2$}, \mbox{-$\frac {1}2$},0,0 \rangle_{\widetilde{R}~}$
}
}
\\ \hline \hline 
single twisted states 
\\ \hline \hline
chiral 
\\ 
\tabu{p{7cm}cp{7cm}}{
\tabu{p{2.5cm}p{4.5cm}}{
$Z^{\rep{T_1}_{[V_4]}} ~:$ 
&
$ | \rep{T_1}_{[V_4]} \rangle \otimes 
| 0, \frac 12, \frac 14, \mbox{-}\frac 14 \rangle_{\widetilde{NS}} $
}
& &
\tabu{p{1.5cm}p{5.5cm}}{
$\gf^{\rep{T_1}_{[V_4]}} ~:$ 
&
$ | \rep{T_1}_{[V_1]} \rangle \otimes   
 |  \frac 12, 0, \mbox{-}\frac 14, \mbox{-}\frac 14 \rangle_{\widetilde{R}~}$
}
}
\\ \hline 
}
\end{center}
\caption{
The local four dimensional spectrum at the fixed points of
$T^6/\Intr_4$ of the heterotic string is given in terms of oscillators
and vacua. The following indices are used: light--cone: 
$\gm,\gn = 2,\dots 3$, internal $T^6$: 
$1, j = 2, 3; \ubar1, \ubar j =  \ubar 2, \ubar 3$, and spinor: $\ga,\gb=\pm$. 
}
\labl{tb:4Dspec}
\end{table}

\begin{table}
\begin{center}
\renewcommand{\arraystretch}{1.1}
\begin{tabular}{|c|c|c|c|c|} \hline
No. & shift & gauge group & \multicolumn{2}{|c|} {untwisted matter}
\\
$8 V_4^2$  & $4V_4$ & $\rep{Ad}_{[V_4]}$ & $\rep{r}_{[V_4]}$ & 
$\rep{R}_{[V_4]}$ 
 \\
\hline \hline 
0 & (00000000) & $\E{8}$  & nothing & nothing 
\\ \hline 
1 & (11000000) & $\E{7} \times \U{1}$  & $(\rep{56})_1$      & 
$(\rep{1})_{2}  + (\rep{1})_{\mbox{-}2} $
\\ \hline
2 & (20000000) & $\SO{14} \times \U{1}$ & $(\rep{64}_s)_1$ &
$(\rep{14}_v)_2  + (\rep{14}_v)_{\mbox{-}2}$ 
\\ \hline
3 & (21100000) & $\E{6}\times \SU{2} \times \U{1}$ & 
$(\crep{27},\rep{2})_1+(\rep{1},\rep{2})_{\mbox{-}3}$ &  
$(\crep{27},\rep{1})_{\mbox{-}2}+(\rep{27},\rep{1})_2$  
\\ \hline
$4_0$ & (22000000) & $\E{7} \times \SU{2}$ & nothing & 
$(\rep{56},\rep{2})$  
\\ \hline
$4_1$ & $(1111111\mbox{-}\!1)$ & $\SU{8} \times \U{1}$ & 
$(\crep{56})_1 + (\rep{8})_{\mbox{-}3}$ & 
$(\rep{28})_2 + (\crep{28})_{\mbox{-}2}$  
\\ \hline 
5 & (31000000) & $\SO{12} \times \SU{2} \times \U{1}$ & 
$(\rep{32}_s,\rep{1})_{\mbox{-}1} + (\rep{12}_v,\rep{2})_1$ & 
$(\rep{32}_c,\rep{1})_0 + (\rep{1},\rep{1})_2
+ (\rep{1},\rep{1})_{\mbox{-}2}$ 
\\ \hline
6 & (22200000) & $\SO{10} \times \SU{4}$ &
$(\rep{16}_c,\crep{4})$ & $(\rep{10}_v,\rep{6})$ 
\\ \hline
7 & (31111100) & $\SU{8} \times \SU{2} $ & 
$(\crep{28},\rep{2})$ & $(\rep{70},\rep{1})$
\\ \hline
8 & (40000000) & $\SO{16}$  &  nothing & $\rep{128}_c$  
\\ \hline
\end{tabular}
\end{center}
\caption{
The resulting gauge groups and six dimensional untwisted
matter representations are given for representatives of the possible
$\Intr_4$ gauge shifts. General gauge shifts, which have been
brought to their standard form, are classified by computing $8V_4^2$. 
For $8V_4^2 = 4$ there are two inequivalent gauge shifts that can be
distinguished by $\sum_I V_4^I \mod 2 = 0,1$, as is discussed in
appendix \ref{sc:Z2class}.   
}
\label{tb:class-4-U}
\end{table}

After the discussion of the local spectra at the six dimensional fixed
hyper surfaces of $T^6/\Intr_4$, the next task is to determine the
four dimensional spectra at the fixed points $\fZ^4_{p\, q}$ of
$T^6/\Intr_4$. Our strategy will be essentially the same as before: we
consider equivalent pure orbifold models, and use those to infer what
the local spectra are in models with Wilson lines. However,
there are a couple of additional complications now: We have to
consider three sectors: ten and six dimensional states, which
are projected at these fixed points, and genuine four dimensional
twisted states. We would like to identify those four dimensional
twisted states using the conditions for massless zero modes on
$T^6/\Intr_4$. It is well--know that for a $\Intr_4$ there are both
single and double twisted zero modes.

These double twisted zero modes are really obtained by
compactification of the six dimensional twisted states to four
dimensions.  But as the
geometrical analysis of $T^6/\Intr_4$ in section \ref{sc:Z4geom}
taught us, $T^6/\Intr_4$ contains six two--tori and four orbifolds
($T^2/\Intr_2$'s), and only these orbifolds contain (as their fixed
points) the fixed points of $T^6/\Intr_4$. Since we are interested in
the local spectra at these fixed points, and we want to use the four
dimensional zero modes to infer these spectra, we have to take care, 
that we do not count zero modes coming from the two--tori, as they
do not live at those four dimensional fixed points. 
But there is an easy way to take this into account. 
Compactifications on $T^2$ and $T^2/\Intr_2$ are closely related: If
the zero mode spectrum on $T^2/\Intr_2$ is in representation
$\rep{T_2}$, the zero mode spectrum on $T^2$ is 
$\rep{T_2} + \crep{T_2}$. 
As argued in section \ref{sc:Z4geom} the orbifold $T^6/\Intr_4$
contains six $T^2$ and four $T^2/\Intr_2$, hence we find the zero
spectrum  $10 \rep{T_2} + 6 \crep{T_2}$. This pattern has been found
for $\Intr_4$ orbifold models in refs.\ \cite{
Katsuki:1989qz,Kawamura:1996zu}. 
At the fixed points of $T^2/\Intr_2$, 
which coincide with four fixed points of $T^6/\Intr_4$, the double
twisted states fall into representations $\rep{T_2}$. Finally,
these double twisted states are the local projections of six
dimensional states, which live in the representation defined in
\eqref{RepCond}: The double twisted representation $\rep{T_2}$ is
obtained by the decomposition 
\equ{
\rep{S}_{[2V_4]} + 4 \rep{D}_{[2V_4]} \ra 
\rep{T_2}_{[V_4]} + \crep{T_2}_{[V_4]}, 
}
associated with the branching $G_{[2V_4]} \ra G_{[V_4]}$ of the six
dimensional gauge group to the four dimensional one. We use $V_4$ to
denote a generic $\Intr_4$ gauge shift vector. 
The possible double twisted representations have been listed table
\ref{tb:class-4-T}.

\begin{table}
\renewcommand{\arraystretch}{1.1}
\begin{tabular}{|c|l|c|c|c|}\hline
shift & \multicolumn{1}{|c|}{gauge group} & single twisted & double twisted  
\\
$V_4$ &  \multicolumn{1}{|c|}{$G_{[V_4]}$}  & $\rep{T_1}_{[V_4]}$ & $\rep{T_2}_{[V_4]}$ 
\\ \hline \hline 
$(3;0)$ & $\E{6} \times \SU{2} \times \U{1}  \times $ & 
$(\crep{27},\rep{1})_{\mbox{-}1/2}(\rep{1})'
+2(\rep{1},\rep{2})_{\mbox{-}3/2}(\rep{1})'$ & 
$(\crep{27},\rep{1})_1(\rep{1})'  +(\rep{1},\rep{1})_{\mbox{-}3}(\rep{1})'$ \\
 & \multicolumn{1}{|r|}{$\E{8}'$} & $+5(\rep{1},\rep{1})_{3/2}(\rep{1})'$ & 
 $ +2(\rep{1},\rep{2})_0(\rep{1})'$ 
\\ \hline
$(3;4_0)$ & $\E{6}\times \SU{2} \times \U{1}  \times $ & 
$(\rep{1},\rep{2})_{\mbox{-}3/2}(\rep{1},\rep{2})'
+2(\rep{1},\rep{1})_{3/2}(\rep{1},\rep{2})'$ &
$(\rep{27},\rep{1})_{\mbox{-}1}(\rep{1},\rep{1})'  
+(\rep{1},\rep{1})_3(\rep{1},\rep{1})'$ \\
 & \multicolumn{1}{|r|}{$\E{7}' \times \SU{2}'$} & 
& $+2(\rep{1},\rep{2})_0(\rep{1},\rep{1})'$
\\ \hline
$(3;4_1)$ & $\E{6} \times \SU{2} \times \U{1}  \times $ & 
$(\rep{1},\rep{1})_{3/2}(\rep{8})_{\mbox{-}1}'
+(\rep{1},\rep{2})_{\mbox{-}3/2}(\rep{1})_{2}'$ & 
$(\rep{1},\rep{2})_{0}(\crep{8})_{\mbox{-}1}'$ 
\\
 & \multicolumn{1}{|r|}{$\SU{8}'  \times \U{1}'$} & 
$+2(\rep{1},\rep{1})_{3/2}(\rep{1})_{2}'$ &
\\ \hline
$(3;8)$ & $\E{6}\times \SU{2} \times \U{1}  \times $ & 
$(\rep{1},\rep{1})_{3/2}(\rep{16}_v)'$ & 
$(\crep{27},\rep{1})_1(\rep{1})'  +(\rep{1},\rep{1})_{\mbox{-}3}(\rep{1})' $ \\
 &  \multicolumn{1}{|r|}{$\SO{16}'$} &  & $+2(\rep{1},\rep{2})_0(\rep{1})'$
\\ \hline
$(7;0)$ & $\SU{8} \times \SU{2} \times  $ &  
$(\crep{8},\rep{2})(\rep{1})'+2(\rep{8},\rep{1})(\rep{1})'$ & 
$(\rep{28},\rep{1})(\rep{1})' +2(\rep{1},\rep{2})(\rep{1})'$ \\
&  \multicolumn{1}{|r|}{$\qquad \times \E{8}'$} & & 
\\ \hline
$(7;4_0)$ & $\SU{8} \times \SU{2} \times $ & 
$(\rep{8},\rep{1})(\rep{1},\rep{2})'$ & 
$(\crep{28},\rep{1})(\rep{1},\rep{1})' 
+ 2(\rep{1},\rep{2})(\rep{1},\rep{1})'$ \\
&  \multicolumn{1}{|r|}{$\E{7}' \times \SU{2}'$} & & 
\\ \hline
$(7;4_1)$ & $\SU{8} \times \SU{2} \times $ & 
$(\rep{8},\rep{1})(\rep{1})_2'$ & $(\rep{1},\rep{2})(\rep{8})_1'$ \\
&  \multicolumn{1}{|r|}{$ \SU{8}' \times \U{1}'$} & & 
\\ \hline
$(7;8)$ & $\SU{8} \times \SU{2} \times  $ & nothing 
 & $(\rep{28},\rep{1})(\rep{1})' +  2(\rep{1},\rep{2})(\rep{1})'$  \\
&  \multicolumn{1}{|r|}{$\SO{16}'$} & & 
\\ \hline
$(2;1)$ & $\SO{14} \times \U{1}  \times $ & 
$(\rep{14}_v)_{\mbox{-}1}(\rep{1})_{1/2}'
+(\rep{1})_{1}(\rep{1})_{\mbox{-}3/2}'$ & 
$(\rep{14}_v)_{0}(\rep{1})_{1}'  + (\rep{1})_{2}(\rep{1})_{\mbox{-}1}' $ \\ 
 &  \multicolumn{1}{|r|}{$\E{7}' \times \U{1}'$} & 
$+5(\rep{1})_{1}(\rep{1})_{1/2}'$ & 
$+ (\rep{1})_{\mbox{-}2}(\rep{1})_{\mbox{-}1}'$ 
\\ \hline
$(2;5)$ & $\SO{14} \times \U{1} \times $ & 
$(\rep{1})_{1}(\rep{12}_v,\rep{1})_{1/2}
+2(\rep{1})_{1}(\rep{1},\rep{2})_{\mbox{-}1/2}$ & 
$(\rep{14}_v)_{0}(\rep{1},\rep{1})_{1}  
+(\rep{1})_{2}(\rep{1},\rep{1})_{\mbox{-}1}  $ \\
 & \multicolumn{1}{|r|}{$\SO{12}' \!\!\times\! \SU{2}' \!\!\times\! \U{1}'$}  & 
&  $+(\rep{1})_{\mbox{-}2}(\rep{1},\rep{1})_{\mbox{-}1}$
\\ \hline
$(6;1)$ & $\SO{10}\times \SU{4} \times $ & 
$(\rep{16}_c,\rep{1})(\rep{1})_{1/2}'
+2(\rep{1},\rep{4})(\rep{1})_{1/2}'$ & 
$(\rep{10}_v,\rep{1})(\rep{1})_{\mbox{-}1}' 
+(\rep{1},\rep{6})(\rep{1})_1'$ \\ 
& \multicolumn{1}{|r|}{$\E{7}' \times \U{1}'$} & & 
\\ \hline
$(6;5)$ & $\SO{10}\times \SU{4}\times $ & 
$(\rep{1},\rep{4})(\rep{1},\rep{2})_{\mbox{-}1/2}'$ & 
$(\rep{10}_v,\rep{1})(\rep{1},\rep{1})_{\mbox{-}1}'
+(\rep{1},\rep{6})(\rep{1},\rep{1})_1'$ \\
&  \multicolumn{1}{|r|}{$\SO{12}'\!\!\times\! \SU{2}' \!\!\times\! \U{1}'$} & & 
\\ \hline
\end{tabular}
\caption{
There are 12 modular invariant combinations of $\Intr_4$
gauge shifts, which are listed in table \ref{tb:class-4-U}. (The numbers 
$(n; n')$ correspond to the first column of that table.) 
The resulting gauge group and the single and double twisted matter at a
single fixed point is given.
}
\label{tb:class-4-T}
\end{table}

Since apart from the double twisted states at the fixed points, the method
working with an equivalent model proceeds as discussed in ref.\
\cite{Gmeiner:2002es}, we only quote our definitions here and give the
spectra in similar tables as tables \ref{tb:class-4-U} and \ref{tb:class-4-T}.
Unlike the discussion of the previous section, here we do not
include any gravitational induced states. The main reason for this is, that in the
end we are interested in (local) anomalies, but gravitational states do
not give rise to anomalies in four dimensions. Hence we can safely
ignore them here. The full local four dimensional spectrum  has been
collected in table \ref{tb:4Dspec}. The possible different gauge
shifts in a single $\E{8}$ are listed in table \ref{tb:class-4-U}, and
table \ref{tb:class-4-T} gives the modular invariant combinations.

The (gauge part of the) untwisted spectrum falls into three categories, which
we can describe using similar notation as in the six dimensional case:
there
is a four dimensional gauge multiplet in the adjoint $\rep{Ad}_{[V_4]}$
(corresponding to the gauge group $G_{[V_4]}$),
and a single chiral multiplet in representation $\rep{R}_{[V_4]}$. 
As can be seen from table \ref{tb:class-4-U} the latter representation
is never complex. Additionally, one encounters two chiral
multiplets in the representation  
\equ{
\rep{r}_{[V_4]} = \{w ~|~ V_4^I w_I = \mbox{$\frac 34$} \mod 1\}. 
\labl{defr}
}
The relevant four dimensional $N=1$ super multiplets for the untwisted
sector have been given in table \ref{tb:4Dspec}. 
The gauge group $G_{[V_4]}$ may contain $\U{1}$ factor(s). 
The generators of these $\U{1}$'s are proportional to the gauge
shift embedded in the Cartan subalgebra: 
\equ{
q_{[V_4]} = V_4^I \, H_I, 
\qquad 
q_{[V_4]}' =  V_4^I \, H_I'
\labl{defqq'}
}
 in the Cartan subalgebra of both $\E{8}$'s.   
These $\U{1}$'s are normalized such that the smallest $\U{1}$ charge 
appearing in the untwisted sectors $\rep{R}_{[V_4]}$ and
$\rep{r}_{[V_4]}$ has absolute value 1. The $\U{1}$ charges of the
untwisted states can be found in table \ref{tb:class-4-U}.

To complete the local four dimensional spectrum, we mention the single
twisted states. Here we can follow the same analysis as in section 
\ref{sc:6Dlocal}: again we find $\smash{N_R^{(1)} = 0}$ leading to 
$\go + \gf = (0, \pm 2, \pm 1, \pm 1)/4$ since $c_{(1)} = 5/16$. 
Invoking the GSO projection and requiring a phase $i$ under the
twist, gives the bosonic vacuum state 
$| 0, \frac 12, \frac 14, -\frac 14 \rangle_{\widetilde{NS}}$. 
For the left--moving sector we find the gauge representations
\equ{
\rep{T_1}_{[V_4]} = 
\bigl\{ w ~\big| ~ 
\mbox{$\frac 12$} (w - V_4)^2 = 
\mbox{$\frac {11}{16}$} - N_L^{(1)} \bigr\}, 
\qquad 
N_L^{(1)} = 0, \mbox{$\frac 14$}, \mbox{$\frac 12$}. 
}
In addition to the single twisted states there are the triple twisted
states. However, it is not hard to show, that they have opposite
chirality and are in the complex conjugate representation
$\crep{T_1}$. This just means, that the single and triple twisted states
combine into chiral multiplets, as given in table \ref{tb:4Dspec}. 
The possible representations $\rep{T}_1$ of within the 12 different
orbifold models are collected in table \ref{tb:class-4-T}.

\section{Orbifold traces and anomaly calculations}
\labl{sc:traces}

In the previous section we saw, that the ten dimensional heterotic
string states are projected at the six and four dimensional fixed
surfaces of $T^6/\Intr_4$. Similarly, the six dimensional states at
the orbifolds $T^2/\Intr_2$ within $T^6/\Intr_4$ give rise to
projected states at the four dimensional fixed points. In particular,
we have computed the representations in which all these states reside. 
To determine these spectra we used equivalent pure orbifold
models. In particular, in section \ref{sc:6Dlocal} we used pure
$T^4/\Intr_2$ orbifold models to determine the local six dimensional
spectra on $T^6/\Intr_4$. However, as we warned at the beginning of
section \ref{sc:6Dlocal} this
method does not directly apply to the computation of anomalies and the
computation of traces of operators on these orbifolds in general. The
task of this section is to obtain exact expressions for such traces on 
the orbifolds $T^6/\Intr_4$ and $T^4/\Intr_2$. Moreover, since some of
the six dimensional twisted states within $T^6/\Intr_4$ live on
orbifolds $T^2/\Intr_2$, we also give a trace formula for that
case. The general machinery for such calculations has been
collected in ref.\ \cite{GrootNibbelink:2003gd} on which this section
is based heavily. We first give the general trace formulae, and
we apply them to compute anomalies in subsection \ref{sc:TrAnoms}.

\subsection{General orbifold trace formulae}
\labl{sc:GenTr}

Example 4.2 of ref.\  \cite{GrootNibbelink:2003gd} gives the
expression for the trace 
$ \Tr_{\Real^4 \!\times\!T^6/\Intr_4,\cR}[\cO]$ 
over the $T^6/\Intr_4$ orbifold Hilbert space of an arbitrary operator
$\cO(x,z;\der)$. The orbifold twist operator 
$\cR = R\, \exp(2\pi i\,\gf^i S_i)$ acts on both, gauge and spacetime
indices. Hence here we may simply copy the result: 
\equ{
\Tr_{\Real^4 \!\times\! T^6/\Intr_4, \cR}\, \Bigl[ \cO \Bigr] = ~
\frac 14 \Tr_{\Real^4\!\times\! T^6/\Intr_4} \Bigl[ \cO \Bigr]
+ 
\frac 14 \cdot \frac 1{16} \sum_{p,q} 
\Tr_{(\Real^4,\fZ^4_{p,q})} \Bigl[
 \cR^{4}_{p,q} \cO_\gTh + (\cR^{4}_{p,q})^3 \cO_{\gTh^3}
\Bigr] + 
\non \\
+ \frac 14 \cdot \frac 1{16}  
 \sum_{p\neq q} 
\Tr_{(\Real^4 \!\times\! T^2, \fZ^2_{p\neq q})}
\Bigl[ \cR^{2}_{p,q} \cO_{\gTh^2} \Bigr]
+ 
\frac 14 \cdot \frac 1{16}
\sum_{p=q} 
\Tr_{(\Real^4 \!\times\! T^2/\Intr_2, \fZ^2_{p=q})}
\Bigl[ \cR^{2}_{p,q} \cO_{\gTh^2} \Bigr],
\labl{Tr10D}
}
where the operators $\cO_{\gTh^i},~ i = 1,2,3$ are defined by 
\equ{
\arry{c}{
\cO_{\gTh} = 
\cO\Bigl(x,z;  \frac 12\der_1, \frac{1+i}2\der_2, \frac{1+i}2\der_3
\Bigr),
\quad 
\cO_{\gTh^2} = 
\cO\Bigr(x,z; 
\der_1, \frac{1}2\der_2, \frac{1}2\der_3
\Bigr), 
\\[2ex]
\cO_{\gTh^3} = 
\cO\Bigl(x,z; 
\frac 12\der_1, \frac{1-i}2\der_2, \frac{1-i}2\der_3
\Bigr)
}
\labl{rescOp}
}
From the first term we learn, that one quarter of the states on
$\Real^4 \times T^6/\Intr_4$ behaves like ten dimensional ones, 
but without any orbifold twist acting on their spacetime or gauge indices.
The second term on the first line of \eqref{Tr10D} is evaluated at the
four dimensional fixed points, and the terms on the second line give
the contributions at the fixed $T^2$'s and $T^2/\Intr_2$'s. 
The presence of Wilson lines in the form of periodicity
matrices $T$ are taken into account by the local matrices 
\equ{
\cR^{4}_{p,q} = e^{2 \pi i\, (v^{4\ I}_{p\, q} H_I + \gf^i S_i)},
\qquad 
\cR^{2}_{p,q} = e^{2 \pi i\, (v^{2\ I}_{p\, q} H_I + 2\,\gf^i S_i)},
\labl{phases}
} 
where $S_i$ denotes the spins in light--cone gauge and the local
shifts are given in \eqref{LocalShifts}. 
It should be stressed, that these traces on the six and four dimensional
fixed surfaces are still taken over the ten dimensional gauge and
spacetime representation space. To obtain traces over the gauge and
spacetime representations of lower dimensional fixed spaces, one needs
to take into account the phase factors coming from \eqref{phases}. 
In the discussion of the application of these formulae to anomalies in
section \ref{sc:TrAnoms},  we explain how this works.

We have seen in section \ref{sc:6Dlocal} that the local six
dimensional spectra of $T^6/\Intr_4$ and $T^4/\Intr_2$ can be
identified locally. However, when one computes traces over the Hilbert
space of $T^4/\Intr_2$ one obtains a somewhat different result than
the one for $T^6/\Intr_4$ given above: 
\equ{
\Tr_{\Real^4 \!\times\! T^2\!\times \! T^2/\Intr_2, \cR}\, 
\Bigl[ \cO \Bigr] = ~
\frac 12 \Tr_{\Real^4\!\times \!T^2\!\times\! T^2/\Intr_2} 
\Bigl[ \cO \Bigr]
+ 
\frac 12 \cdot \frac 1{16} \sum_{p,q} 
\Tr_{(\Real^4 \!\times\! T^2,\fZ^2_{p,q})} 
\Bigl[ \cR^{2}_{p,q} \cO_{\gTh^2} \Bigr],
\labl{Tr10DZ2}
}
here $\cO_{\gTh^2}$ is again given by \eqref{rescOp}. 
(This result is obtained by the methods discussed in 
ref.\ \cite{GrootNibbelink:2003gd}.) The two main differences with 
\eqref{Tr10D} are, that the traces at the four dimensional fixed
points are absent, of course, and that there are now factors of $1/2$
instead of $1/4$.

As the discussion in \mbox{section \ref{sc:Z4geom}} revealed, the orbifold
$T^6/\Intr_4$ contains fixed orbifolds $T^2/\Intr_2$, which 
support six dimensional twisted states as given in section
\ref{sc:6Dlocal}, therefore we also give the final trace formula for
this case. The trace formula at such an 
orbifold $(\Real^4\!\times\! T^2/\Intr_2, \fZ^2_{p=q})$ for an
operator $\smash{\tilde\cO}$ that acts on these twisted states reads:
\equ{
\Tr_{(\Real^4 \!\times\! T^2/\Intr_2,\fZ^2_{p=q}), \cR}\, 
\Bigl[ \tilde \cO \Bigr] = ~
\frac 12 \Tr_{(\Real^4\!\times \! T^2/\Intr_2, \fZ^2_{p=q})} 
\Bigl[ \tilde \cO \Bigr]
+ 
\frac 12 \cdot \frac 1{4} \sum_{p_1,q_1} 
\Tr_{(\Real^4,R_1\gz_{p_1q_1},\fZ^2_{p=q})} \Bigl[
 \cR^{4}_{p=q} \tilde \cO_\gTh 
\Bigr],
\labl{Tr6D}
}
where $\tilde \cO_\gTh = \cO(x,z; \frac 12\der_1)$. 
The pattern of this expression is similar to \eqref{Tr10DZ2}: There is a
six dimensional part, where the trace is evaluated on the orbifold
$T^2/\Intr_2$ without any orbifold twist in the spacetime or gauge
sector. And in addition, for the second term we can recognize the
factor $1/4$ that arises because $T^2/\Intr_2$ has four fixed
points. (The expression here is consistent, because for $p_2=q_2$ and
$p_3=q_3$ the matrix $\cR^{4}_{p,q}$ squares to the identity.)

\subsection{Anomaly calculations using trace formulae}
\labl{sc:TrAnoms}

We now briefly describe how these general trace formulae for
$T^6/\Intr_4$, $T^4/\Intr_2$ and $T^2/\Intr_2$ can be applied to the
evaluation of anomalies. Anomalies correspond to the formal Hilbert
space trace in $2N$ dimensions  
\equ{
\cA_{2N}(\gL) = 2 \pi i\, \Tr( \tgG \gL),
}
where $\gL$ corresponds to a gauge or local Lorentz transformation,
and $\tgG$ is the chirality operator. To evaluate this trace one needs
to regularize this expression, for example, by employing the heat kernel
regularization following Fujikawa
\cite{Fujikawa:1979ay,Fujikawa:1980eg}. 
As the treatment of this method to orbifolds has been recently
discussed in refs.\ 
\cite{Gmeiner:2002es,Asaka:2002my,vonGersdorff:2003dt}, 
we here only focus on some important consequences that can be read off
from the trace formulae given in the previous subsection.

Let us start with applying the result \eqref{Tr10D} for $T^6/\Intr_4$
to anomalies. We see that there is a ten dimensional anomaly but with
a relative normalization factor of $1/4$ w.r.t.\ the result on a
smooth ten dimensional manifold. Next, the two
expressions on the second line refer to trace contributions at the six
dimensional fixed hyper surfaces, $\Real^4 \times T^2$ and
$\Real^4 \times T^2/\Intr_2$ of $T^6/\Intr_4$, respectively. As
stressed below that equation, the traces are still taken over gauge and
spacetime representations in ten dimensions, while we would like to
express the result for the anomaly in six dimensional representations
only. To do this we need to keep track of the phase factors resulting
from $\cR^2_{p, q}$, given in \eqref{phases}. We will not give this
calculation for all states, but just illustrate the method for the ten
dimensional gaugino. This state can be decomposed as 
\equ{
\arry{c}{
\Bigl( 
| \rep{Ad}_{[v^2_{p\,q}]} \rangle +  | \rep{R}_{[v^2_{p\,q}]} \rangle 
\Bigr)
| \frac {\ga_2\ga_3\ga_4}{2}, \frac {\ga_2}{2}, 
\frac {\ga_3}{2}, \frac {\ga_4}{2} \rangle_{\widetilde{NS}},
}
}
where we have taken into account that the ten--dimensional gaugino is
left--handed. The six dimensional chirality of these states is given
by $\ga_1\ga_2 = \ga_3\ga_4$, and their phase factors are computed
easily: 
\equ{
\arry{l}{
\cR^2_{p, q}\,  | \rep{Ad}_{[v^2_{p\,q}]} \rangle 
| \frac {\ga_2\ga_3\ga_4}{2}, \frac {\ga_2}{2}, 
\frac {\ga_3}{2}, \frac {\ga_4}{2} \rangle_{\widetilde{NS}} = 
e^{2\gp i\, \bigl[ 0 + \frac 14( \ga_3 + \ga_4 -2 \ga_2)\bigr]} \, 
| \rep{Ad}_{[v^2_{p\,q}]} \rangle 
| \frac {\ga_2\ga_3\ga_4}{2}, \frac {\ga_2}{2}, 
\frac {\ga_3}{2}, \frac {\ga_4}{2} \rangle_{\widetilde{NS}}, 
\\[2ex]
\cR^2_{p, q}\,  | \rep{R}_{[v^2_{p\,q}]} \rangle 
| \frac {\ga_2\ga_3\ga_4}{2}, \frac {\ga_2}{2}, 
\frac {\ga_3}{2}, \frac {\ga_4}{2} \rangle_{\widetilde{NS}} \ \ = 
e^{2\gp i\, \bigl[ \frac 12 + \frac 14( \ga_3 + \ga_4 -2 \ga_2)\bigr]} \, 
| \rep{R}_{[v^2_{p\,q}]} \rangle 
| \frac {\ga_2\ga_3\ga_4}{2}, \frac {\ga_2}{2}, 
\frac {\ga_3}{2}, \frac {\ga_4}{2} \rangle_{\widetilde{NS}}. 
}
}
The factor $\exp 2\pi i\, \frac 12$ leads to the opposite sign between
the states in $\rep{Ad}_{[v^2_{p\,q}]}$ and
$\rep{R}_{[v^2_{p\,q}]}$. This reflects the opposite six dimensional
chirality of vector and hyper multiplets. Computing the remaining sum,
noting that $\exp 2\pi i\, \frac 12 \ga_2 = -1$ gives
\equ{
\sum_{\ga_3,\ga_4} (\ga_3\ga_4) 
e^{2\gp i\, \bigl[\frac 14( \ga_3 + \ga_4 -2\ga_2)\bigr]}
= 4, 
\labl{symFac10D}
}
shows that the factor $1/4$ of \eqref{Tr10D} is canceled. Notice that
this leaves only $\ga_2 = \pm$ to give two degrees of freedom. However,
since six dimensional chiral spinors contain four degrees of freedom
(on-shell) we have to introduce a factor $1/2$ to normalize the
anomaly to the anomaly of a chiral six dimensional fermion; this
result is given in equation \eqref{10DgauginoAnom} below.

Similar arguments can be used to evaluate the anomaly at the four
dimensional fixed points. From the second (two) terms of \eqref{Tr10D} we
obtain the symmetrization factors due to  
$\cR^4_{p\, q}$ and $(\cR^4_{p\,q})^3 = (\cR^4_{p\,q})\inv$:
\equ{
\sum_{\ga_3,\ga_4 = \pm} 
\Bigl\{
e^{2\pi i\, \bigl[ \frac 34 + \frac 18( \ga_3 + \ga_4 - 2 \ga_3\ga_4)\bigr]}
+ 
e^{-2\pi i\, \bigl[ \frac 34 + \frac 18( \ga_3 + \ga_4 - 2 \ga_3\ga_4)\bigr]}
\Bigr\} = 4, 
}
since all ten dimensional states 
\(
| \rep{r}_{[v^4_{p\,q}]} \rangle \otimes 
| \frac 12, \frac {\ga_3\ga_4}2, \frac {\ga_3}2,\frac {\ga_4}2 
\rangle_{\widetilde{NS}}
\) 
in this representation contribute (only the four dimensional chirality
has already been fixed). This cancels one $1/4$, and the factor $1/16$
remains. The other four dimensional representations
$\rep{Ad}_{[v^4_{p\,q}]}$ and $\rep{R}_{[v^4_{p\,q}]}$ are not
complex, and hence do not give rise to a four dimensional
anomaly.

Putting all this together, we find that the final result for the
gaugino anomaly is given by  
\equa{
\cA_{\Real^4\!\times \! T^6/\Intr_4|\E{8}^2}(\gL) = & 
\int  \Bigl\{
\frac 14 \cdot \frac 12 \, I^1_{10|\rep{Ad}_{[0]}}
+ 
\sum_{p,q}
\frac 1{16} \cdot \frac 12\, 
\bigl( I^1_{6|\rep{Ad}_{[v^2_{p\,q}]}} - 
I^1_{6|\rep{R}_{[v^2_{p\,q}]}} \bigr)
\gd^4(z - \fZ^2_{p\, q}) \d^4 z 
\non \\ 
& + 
\sum_{p,q}
\frac 1{16} \cdot 2 \,  I^1_{4|\rep{r}_{[v^4_{p\,q}]}} 
\gd^6(z - \fZ^4_{p\, q}) \d^6 z 
\Bigr\}. 
\labl{10DgauginoAnom}
}
Here $I^1_{2N}$ denote the anomalies associated with the closed and 
gauge invariant anomaly polynomials $I_{2N+2}$ via the descent
equations 
\equ{
\d I_{2N+1} = I_{2N +2}, 
\quad 
\gd_\gL I_{2N+1} = \d \, I_{2N}^1; 
\qquad \text{with} ~~
I_{2N+2|\rep{r}} = \hat A(R) \text{ch}_{\rep{r}}(iF), 
}
where in the last equation the general ($2N+1$)--form anomaly
polynomial for a chiral fermion in representation $\rep{r}$ is
defined. Here $\hat A(R)$ is the roof genus of the curvature two form
$R$, and $\text{ch}_{\rep{r}}(iF)$ the Chern character of the field
strength two form $F$, for their definitions see e.g.
\cite{Zumino:1984rz,Bardeen:1984pm,Alvarez-Gaume:1984ig,Nakahara:1990th}.
(Throughout this work we assume wedge products between forms to be
understood.) In ten dimensions, spinors are both chiral and Majorana,
this means that only half of the states contribute to the
anomalies; this explains the appearance of the factor $1/2$ of the
first term of \eqref{10DgauginoAnom}. As discussed above, the factor
$1/2$ in front of the second and third terms is due to the normalization
of chiral anomalies in six dimensions. The factor $2$ in
front of the last term  is due to the fact that states in
representation $\rep{r}_{[V_4]}$ have multiplicity 2, see table
\ref{tb:4Dspec}.

It should be noted that the calculation presented in
\eqref{symFac10D}, where all traces have been expressed in a six dimensional form,
applies equally well to the case of the orbifold $T^4/\Intr_2$. Using 
the trace formula \eqref{Tr10DZ2} for that orbifold we find for the
gaugino anomaly 
\equ{
\cA_{\Real^4\!\times \! T^2\!\times\!T^4/\Intr_2|\E{8}^2}(\gL) =  
\int  \Bigl\{
\frac 12 \cdot \frac 12 \, I^1_{10|\rep{Ad}_{[0]}}
+ 
\sum_{p,q}
\frac 1{16} \, 
\bigl( I^1_{6|\rep{Ad}_{[v^2_{p\,q}]}} -  
I^1_{6|\rep{R}_{[v^2_{p\,q}]}} \bigr)
\gd^4(z - \fZ^2_{p\, q}) \d^4 z \Bigr\}. 
\labl{10DgauginoAnomZ2}
}
Observe that there is a difference of a factor of two between this
expression and \eqref{10DgauginoAnom} for both the ten and six
dimensional anomalies.

The computation of the anomaly for six dimensional states on
$T^2/\Intr_2$ is straightforward and similar to the results presented
above. Using \eqref{Tr6D} we find for the half--hyperinos 
\equ{
\cA_{(\Real^4\!\times \! T^2/\Intr_2, \fZ^2_{p\,q})}(\gL) =  
\int  \Bigl\{
\frac 12 \cdot \frac 12
\bigl( 
I^1_{6|\rep{S}_{v^2_{p\,q}}} +  I^1_{6|\rep{D}_{v^2_{p\,q}}}
\bigr) 
+ 
\sum_{p_1,q_1}
\frac 1{16} \,  
I^1_{4|\rep{T_2}_{[v^4_{p\,q}]}} 
\gd^2(z_1 - R_1\gz_{p\, q}) \d^2 z_1 \Bigr\}. 
\labl{6DAnomZ2}
}
For the hyperinos that live on the two $T^2$'s which are identified,
the anomaly is given by the first two terms in this equation, because
like for the orbifold only half of the states exist at a two--torus.

\section{Anomalies}
\labl{sc:Anoms}

Anomaly investigations of ten dimensional theories have played an
important role in the development of string theory. In particular the 
possibility of canceling factorisable anomalies via the so--called
Green--Schwarz mechanism \cite{Green:1984sg} paved the way for
heterotic string theory. On orbifolds the ten
dimensional factorization is still at work. The 12--form
anomaly polynomial is given by
\equ{
I_{12} = 
\Bigl[ \tr R^2 - \frac 1{30} \tr (iF)^2  - \frac 1{30} \tr (iF')^2 
\Bigr] \, \frac 1{4} X_{8\, GS} \; ,
\labl{eq:fact10danom} 
}
where $X_{8\, GS}$ is the standard Green--Schwarz eight form 
\cite{Green:1984sg,gsw_2,Green:1985bx}. Here we have explicitly given
the factor $1/4$ that arises because we have to evaluate the anomalies
on the orbifold $T^6/\Intr_4$. From the first term of
\eqref{10DgauginoAnom}, we see that we only get  
$1/4$ of the anomaly on a smooth manifold. Next we investigate the
anomalies at the fixed spaces of the orbifold $T^6/\Intr_4$.

\subsection{Local six dimensional anomalies}
\labl{sc:6DAnoms}

Six dimensional anomalies arising from compactification of 
ten dimensional supergravity coupled to $\E{8} \times \E{8}'$ super
Yang--Mills have first been considered in \cite{Green:1985bx}. The 
application to heterotic orbifold models can be found e.g.\ in
\cite{Erler:1994zy}. The relevant representations 
for the local six dimensional anomaly investigation have been given in
section \ref{sc:6Dlocal}. We denote the six dimensional anomaly
polynomial for a fermion with positive chirality transforming in
representation $\rep{r}$ by
\begin{equation}
I_{8|\rep{r}} = \frac{-i}{(2\pi)^3} \Bigl[ 
\frac1{24} \tr_{\rep{r}} (iF)^4 
- \frac1{96} \tr_{\rep{r}} (iF)^2 \tr R^2 
+ \frac{\dim \rep{r}}{128} 
\Bigl( 
\frac1{45} \tr R^4 + \frac1{36} (\tr R^2)^2 
\Bigr) 
\Bigr].
\labl{6DAnomr}
\end{equation}
(This formula also applies to gauge singlets, in which case all traces 
$\tr_{\rep{r}}$ are of course zero and $\dim \rep{r}$ denotes the
number of these gauge singlets. For a single gauge singlet we denote
its anomaly polynomial by $I_{8|1/2}$.) 
This equation directly applies to gauginos, because they have positive
chirality. Since the chirality of the hyperinos and the dilatino
is negative, the overall sign of the anomaly polynomial is
opposite. (We give this sign explicitly to avoid confusion.) In
addition, the anomaly contribution of the gravitino reads  
\begin{equation}
I_{8|3/2} = 
\frac{-i}{(2\pi)^3} \frac 1{128} 
\Bigl(
\frac{245}{45} \tr R^4 - \frac{43}{36} (\tr R^2)^2 
\Bigr).
\end{equation}
Combining these ingredients, the total anomaly polynomial for the
heterotic $T^6/\Intr_4$ orbifold model at a six dimensional
fixed point $\fZ^2_{pq}$ is given by 
\begin{equation}
I_{8|\fZ^2_{pq}} = 
\frac1{32}\, I_{8|3/2} 
-  \frac1{32}(1 + 4)  I_{8|1/2} 
+ \frac1{32}\, I_{8|\rep{Ad}_{[v^2_{p\,q}]}} 
- \frac1{32} I_{8|\rep{R}_{[v^2_{p\,q}]}} 
- \frac14 \Bigl( I_{8|\rep{S}_{[v^2_{p\,q}]}} + I_{8|\rep{D}_{[v^2_{p\,q}]}} \Bigr).
\labl{6DAnom}
\end{equation}
The different contributions are due to the gravitino, the dilatino and
the 4 neutral hyperinos, the gaugino, the untwisted matter, and the
twisted matter, respectively (see table \ref{tb:6Dspec}). 
Here we used the anomaly results \eqref{10DgauginoAnom} and
\eqref{6DAnomZ2} obtained in section \ref{sc:TrAnoms}. 
(Only the gauginos are treated there, but the
discussion extends to gravitino and dilatino as well.)
We should note that if one does this analysis for $T^4/\Intr_2$ one
obtains two times the result of \eqref{6DAnom}, which follows from 
\eqref{10DgauginoAnomZ2} and the fact that in that case the
(half--)hyperinos on the six dimensional spaces are not orbifolded.

\begin{table}
\[
\renewcommand{\arraystretch}{1.1}
\arry{|c||c|cccc|}{
\hline 
\text{model}&G_i & \E{7} & \SU{2} & \E{8}' & \SO{16}' 
\\  \hline \hline
 & c_i & 1/6 &  2 & 1/30 & 1 
\\ \hline 
(1,0) & d_i & 1 &  12 &  \mbox{-}1/5 & \text{--} 
\\ 
(1,2)& d_i & \mbox{-}1/3 &  28 & \text{--} & 2
\\ \hline 
}
\]
\caption{\label{tb:6danomfac}
This table gives the factorization coefficients $c_i$ and $d_i$, defined by 
(\ref{eq:fact6danom}), for the two heterotic models $(1,0)$ and 
$(1,2)$ as given in table \ref{tb:class-2-T}. 
}
\end{table}

Since both six dimensional models we encounter in this work (see table
\ref{tb:class-2-T}) do not contain Abelian subgroups, we are concerned
with semi--simple gauge groups only. The non--Abelian anomalies do not
have to vanish identically, but can be canceled by a six dimensional
Green--Schwarz mechanism instead. For this it is crucial that the
anomaly polynomial factorizes, which means in particular that all 
irreducible anomalies have to vanish identically. For both models the
irreducible gravitational anomaly $\tr R^4$ vanishes because of the
relation between the number of vector and hyper multiplets 
\begin{equation}
\label{eq:factreq6d}
\frac 1{16}\Bigl( 245 - 1 - 4 \Bigr) + 
\frac 1{16} \dim \rep{Ad}_{[v^2_{p\,q}]}
-\frac{1}{16} \dim \rep{R}_{[v^2_{p\,q}]} 
- \frac 12 \Bigl( \dim \rep{S}_{[v^2_{p\,q}]} + \dim \rep{D}_{[v^2_{p\,q}]} \Bigr) 
= 0, 
\end{equation}
at a given six dimensional hyper surface (This is nothing but the local
version of the well--known zero--mode statement, that the total number of hyper--
minus vector multiplets must be $244$). Of the groups appearing in 
table \ref{tb:class-2-T}, only $\SO{16}'$ has a non--vanishing fourth
order Casimir and may therefore lead to irreducible gauge anomalies. 
However, by virtue of the $\SO{16}'$ trace identities
\cite{vanRitbergen:1998pn,Erler:1994zy}  for the adjoint and spinor
representation 
\equ{
\tr_{\rep{120}}(iF)^4 = 8\, \tr_{\rep{16}}(iF)^4 
+ 3\bigl(\tr_{\rep{16}}(iF)^2 \bigr)^2, 
\quad
\tr_{\rep{128}}(iF)^4 = - 8\, \tr_{\rep{16}}(iF)^4 
+ 6\bigl(\tr_{\rep{16}}(iF)^2 \bigr)^2, 
\labl{Tr4SO16}
}
also these irreducible anomaly contributions of the gaugino and the 
untwisted and twisted matter cancel each other:  
$(1/16) 8 - (1/16)(-8) -(1/2) 2 = 0.$ 
The remaining six dimensional reducible anomaly at a given fixed point
$\fZ^2_{pq}$ always factorizes into the form
\equa{
I_{8|\fZ^2_{pq}}  = &
- \Bigl[ \tr R^2 - \sum_i c_i \tr (iF_i)^2 \Bigr]
\frac{-i}{(2 \pi)^3} \frac1{16^2} 
\Bigl[ \tr R^2 - \sum_i d_i \tr (iF_i)^2 \Bigr] 
= \frac{-i}{(2 \pi)^3} \frac1{16^2} \Bigl[ 
- (\tr R^2)^2 + \Bigr.  
\non 
\\[2ex]
& 
\Bigl. + 
\sum_i \Bigl( 
(c_i + d_i) \tr R^2 \tr (iF_i)^2 - c_i d_i ( \tr (iF_i)^2)^2 
\Bigr)
- \sum_{j\neq i} c_i d_j \tr (iF_i)^2 \tr (iF_j)^2
\Bigr],
\label{eq:fact6danom} 
}
where $i,j$ runs over the (semi--simple) gauge group factors and the
traces are taken in the corresponding fundamental representations. 
The factors $c_i$ give universal normalization of the quadratic traces
in the sense,  that they only depend on the gauge groups: 
$c_i = 2/I(G_i)$, where $I(G_i)$ is the index of the group $G_i$. 
On the other hand, the coefficients $d_i$ are model dependent, as 
can be seen from table \ref{tb:6danomfac}, where we have listed $c_i$ and
$d_i$ for both models of table \ref{tb:class-2-T}.

We do not give the details for the derivation of these coefficients from
\eqref{6DAnomr} and \eqref{6DAnom}; they are
obtained using trace identities, like \eqref{Tr4SO16}, which can be
found e.g. in \cite{vanRitbergen:1998pn,Erler:1994zy}. However, we would
like to remark that the relation between these coefficients is rather 
delicate: For example, the mixed gauge anomalies, that appear in the
last term in the second line of \eqref{eq:fact6danom}, can only arise
if there is matter charged under two gauge group factors $G_i$ and
$G_j$ simultaneously. 
Therefore, if two group factors do not have any matter that is charged under
both, their coefficients satisfy $c_i d_j + d_i c_j = 0$. In model
$(1,0)$ there is no mixed matter that is charged under $\E{8}'$, hence 
this relation is fulfilled for $j = \E{8}'$, as is readily verified
from table \ref{tb:6danomfac}. Model $(1,2)$ does not contain any
matter charged under both $\E{7}$ and $\SO{16}'$, so a similar
conclusion holds.

\subsection{Local four dimensional anomalies} 
\labl{sc:4DAnoms}

The relevant representations for the local four dimensional anomaly
investigation have been given in section \ref{sc:4Dlocal}. We denote
the four dimensional anomaly polynomial for a fermion with positive
chirality transforming in representation $\rep{r}$ by 
\equ{
I_{6|\rep{r}} = \frac {-i}{(2\pi)^2} 
\Bigl[
\frac 1{3!} 
\tr_{\rep{r}}(iF)^3 - \frac 1{48} \tr R^2 \tr_{\rep{r}}(iF) 
\Bigr] \; .
}
The general form of the anomaly
polynomial at the four dimensional fixed point $\fZ^4_{p\, q}$ of
$T^6/\Intr_4$ is given by 
\equ{
I_{6| \fZ^4_{p\, q}} = 
\frac 1{16} \cdot 2 \, I_{6|\rep{r}_{[v^4_{p\,q}]}}
+ \frac 1{4} \, I_{6|\rep{T_2}_{[v^4_{p\,q}]}} 
+ I_{6|\rep{T_1}_{[v^4_{p\,q}]}}. 
\labl{4DAnom}
}
This result is obtained by combining \eqref{10DgauginoAnom} and
\eqref{6DAnomZ2}.  The first term is due to the part of the ten
dimensional gauginos that resides in representation $\rep{r}_{[v^4_{p\,q}]}$. 
The other ten dimensional states do not give rise to anomalies since
their representations $\rep{Ad}_{[v^4_{p\,q}]}$ and
$\rep{R}_{[v^4_{p\,q}]}$ are by definition \eqref{defAdR}  never
complex. The six dimensional twisted states at the orbifold
$T^2/\Intr_2$ (that contains $\fZ^4_{p\, q}$) give rise to the second
term in \eqref{4DAnom}.

We show that there are no non--Abelian anomalies at any of the four
dimensional fixed points. The relevant groups and representations for
the local anomaly analysis have been given in tables
\ref{tb:class-4-U} and \ref{tb:class-4-T}. Of all possible local
non--Abelian gauge groups only $\SU{8}$, $\SU{4}$ and $\SU{8}'$ are
not automatically anomaly free. (Here we refer to table
\ref{tb:class-4-T} to identify these groups.) In table 
\ref{tb:4DAnomsNA} we verify explicitly that all models, that contain 
these gauge groups, do not suffer from non--Abelian anomalies. 
For the $\SU{4}$ cases this is straightforward, for some of the
$\SU{8}$ and $\SU{8}'$ cases, trace identities are needed (cf. e.g. 
\cite{vanRitbergen:1998pn,Erler:1994zy}).

\begin{table}
\begin{center}
\renewcommand{\arraystretch}{1.1}
\tabu{|c|c|c|c|c|}{ 
\hline 
group & model & 
\multicolumn{3}{|c|}{local 4D anomaly contributions} 
\\
$G_{[V_4]} \supset$ & $V_4$ & 
$\frac 1{16} \cdot 2\, I_{6|\rep{r}_{[V_4]}}$ & 
$ \frac 1{4} \cdot I_{6|\rep{T_2}_{[V_4]}}$ &
$I_{6|\rep{T_1}_{[V_4]}}$
\\ \hline\hline 
$\SU{4}$ & $(6,1)$; $(6,5)$ & 
$\frac 1{16} \cdot 2\cdot 16\,  I_{6|\crep{4}}$ & 
-- & $ 2 \cdot I_{6|\rep{4}}$
\\ \hline 
$\SU{8}$ & $(7,0)$ & 
$\frac 1{16} \cdot 2\cdot 2 \, I_{6|\crep{28}}$ & 
$\frac 1{4} \cdot  I_{6|\rep{28}}$ & 
$2 \cdot  I_{6|\rep{8}} + 2 \,  I_{6|\crep{8}}$
\\ 
 & $(7,4_0)$ & 
$\frac 1{16} \cdot 2\cdot 2 \,  I_{6|\crep{28}}$ & 
$\frac 1{4} \cdot  I_{6|\crep{28}}$ & 
$2 \cdot  I_{6|\rep{8}}$
\\ 
& $(7,4_1)$ & 
$\frac 1{16} \cdot 2\cdot 2 \,  I_{6|\crep{28}}$ & 
-- & $I_{6|\rep{8}}$
\\ 
& $(7,8)$ & 
$\frac 1{16} \cdot 2\cdot 2 \,  I_{6|\crep{28}}$ & 
$\frac 1{4} \cdot  I_{6|\rep{28}}$ & --
\\ \hline 
$\SU{8}'$ & $(3,4_1)$; $(7,4_1)$ & 
$\frac 1{16} \cdot 2\, (  I_{6|\crep{56}} + I_{6|\rep{8}})$ & 
$\frac 1{4} \cdot  2\,  I_{6|\crep{8}}$ & 
$I_{6|\rep{8}}$
\\ \hline 
}
\end{center}
\caption{
This table shows that all non--Abelian anomalies, which may arise in the
eight models with non--automatically anomaly free groups, are
canceled locally at each four dimensional fixed point of
$T^6/\Intr_4$. 
}
\labl{tb:4DAnomsNA}
\end{table}

Models that contain $\U{1}$ factors may be anomalous, but like in ten
and six dimensions the anomaly polynomial factorizes 
\cite{Schellekens:1987xh}: 
\equ{
I_{6|\fZ^4_{p,q}} = 
\Bigl[ \tr R^2 - \sum_i c_i \tr (iF_i)^2 \Bigr] 
\frac {-i}{(2\pi)^2} \frac 1{48} 
\Bigl[ F_{\U{1}}\, \tr (q_{v^4_{p\,q}}) + F_{\U{1}'}\, \tr (q_{v^4_{p\,q}}')
\Bigr]
\labl{eq:fact4danom}
}
where the coefficients $c_i$ are again related to the indices of the 
various groups that exist at this four dimensional fixed point. 
Because factorization only allows a single field strength 2--form to
appear on the right hand side of \eqref{eq:fact4danom}, only $\U{1}$
factors may have pure and mixed anomalies. Here the charges
$q_{v^4_{p\,q}}$ and $q'_{v^4_{p\,q}}$ are defined as in
\eqref{defqq'}. In fact, it is simply the local sum of 
$\U{1}$ charges that decide whether a given $\U{1}$ factor is
anomalous or not. In table \ref{tb:U1sums} we have computed the sum of
charges for all models with $\U{1}$ factors, listed in table
\ref{tb:class-4-T}. The models $(3;0)$ and $(3;4_0)$ do not have
an anomalous $\U{1}$ even though they contain $\U{1}$ factors. All other
models considered in table \ref{tb:U1sums} have only one anomalous
$\U{1}$ in one of the two $\E{8}$ factors, except for model $(2;5)$,
which seems to have two. However, as observed in ref.\  
\cite{Kobayashi:1997pb} one can always find two other linear
combinations of the charges $q_{v^4_{p\,q}}$ and $q'_{v^4_{p\,q}}$,
such that only one of them is anomalous.

\begin{table}
\[
\renewcommand{\arraystretch}{1.2}
\arry{|c|c|c|c|c|c|c|c|c|c|}{ 
\hline 
\multicolumn{10}{|c|}{\text{models with $\U{1}$ factors}} 
\\
& (3;0) & (3;4_0) & (3;4_1) & (3;8) & (7;4_1) & (2;1) & (2;5) 
& (6;1) & (6;5) 
\\ \hline\hline 
\frac {2}{16} \tr_{\rep{r}_{[V_4]}}(q) & 
6 & 6 & (6,4) & 6 & 4 & (8,7) & (8,\mbox{-}1) & 7 & \mbox{-}1  
\\ 
\frac {1}{4} \tr_{\rep{T_2}_{[V_4]}}(q) & 
6 & \mbox{-}6 & (0,\mbox{-}4) & 6 & 4 & (0,3) & (0,3) &\mbox{-}1 
& \mbox{-}1 
\\ 
\tr_{\rep{T_1}_{[V_4]}}(q) & 
\mbox{-}12 & 0 & (12,0) & 24 & 16 & (\mbox{-}8,8) & 
(16,4) & 12 & \mbox{-}4 
\\ \hline 
\text{sum} & 
0 & 0 & (18,0) & 36 & 24 & (0,18) & (24,6) & 18 & \mbox{-}6
\\ \hline 
}
\]
\caption{
The sum of charges of models with $\U{1}$ factor(s) is computed for 
each matter sector ($\rep{r}_{[V_4]}$, $\rep{T_2}_{[V_4]}$ and 
$\rep{T_1}_{[V_4]}$) separately. (If a model contains two $\U{1}$'s, the
bracket indicate the sum of charges of the $\U{1}$'s in the $\E{8}$
and $\E{8}'$ sectors.) The sum of these three contributions
determines whether that $\U{1}$ is anomalous or not. 
}
\labl{tb:U1sums}
\end{table}

\subsection{Local Green--Schwarz mechanisms} 
\labl{sc:GSlocal}

To conclude the discussion of anomalies, we return to the
Green--Schwarz mechanism to cancel the leftover and factorized
anomalies locally at all the fixed hyper surfaces of $T^6/\Intr_4$. 
Since the essence of this mechanism on orbifolds has
recently been discussed in ref.\ \cite{GrootNibbelink:2003gb},
we here only quote the particularities of the investigation
on $T^6/\Intr_4$.

The theory of $N=1$ supergravity in ten dimensions can be formulated 
using a two form $B_2$ (alternatively one can use a six form potential  
\cite{Chamseddine:1981ez,Bergshoeff:1982um,Chapline:1983ww}) 
described by the action 
\equ{
S_{GS} = \int   - \frac 12 *\d B_2 \, \d B_2 + (*X_3 + X_7 ) \d B_2  
- \frac 12 * X_3 \, X_3, 
\labl{LagrBC}
}
which is invariant under the natural gauge transformations of the 
2--form $\gd_{\gL_1} B_2 = \d \gL_1$. Here the 3-- and 7--forms 
$X_3, X_7$ are derived from arbitrary closed and gauge invariant
4-- and 8--forms, $X_4, X_8$, by Poincar\'e's lemma  
(i.e.\ locally $\d X_3 = X_4$ and $\d X_7 = X_8$). 
The gauge variation of the 2--form and its action reads
\equ{
\gd_\gL B_2 = X_2^1, 
\qquad 
\gd_\gL S_{GS} = \int X_7\, \gd_\gL X_3,
\labl{VarActASTens}
}
where $\gL$ may refer to either a gauge transformation 
$\gd_\gL A_1 = \d \gL + [\gL, A_1]$ or a local Lorentz transformation 
$\gd_L \go_1 = \d L + [L, \go_1]$ of the spin connection one form 
$\go_1$.

In the preceding subsections we have observed that the ten, six and
four dimensional anomalies factorize, see equations 
\eqref{eq:fact10danom}, \eqref{eq:fact6danom} and 
\eqref{eq:fact4danom}. The anomalous variation of the Green--Schwarz
action \eqref{VarActASTens} can cancel all these anomalies
simultaneously, by taking 
\equ{
\arry{lcl}{ \dsp 
X_4 &= & \tr R^2 - \frac 1{30} \tr (iF)^2  - \frac 1{30} \tr (iF')^2 
\\[2ex] \dsp 
X_8 &=& \frac 14\, X_{8\, GS} + 
\frac{-i}{(2 \pi)^3} \frac1{16^2} 
\Bigl[ \tr R^2 - \sum_i d_i \tr (iF_i)^2 \Bigr]_{\fZ^2_{p\,q}} \, 
\gd^4(z - \fZ^2_{p\, q}) \d^4 z 
\\[2ex] \dsp 
& & 
+ \frac {-i}{(2\pi)^2} \frac 1{48} 
\Bigl[ F_{\U{1}}\, \tr (q) + F_{\U{1}'}\, \tr (q')
\Bigr]_{\fZ^4_{p\, q}} \, 
\gd^6(z - \fZ^4_{p\, q}) \d^6 z. 
}
}
Here the notation $[ \ldots ]_{\fZ^2_{p\, q}}$ signifies that the
expression in between the brackets is evaluated at fixed hyper surface
$\fZ^2_{p\, q}$, and so on. The reason that these forms can cancel the
anomalies in the various dimensions is, that when $X_4$ is restricted 
to a lower dimensional hyper surface, the terms 
$(1/30) \tr(iF)^2+(1/30) \tr(iF')^2$ always reduce to 
$\sum_i c_i\, \tr (iF_i)^2$, which appears in the factorizations
\eqref{eq:fact6danom} and  \eqref{eq:fact4danom} in six
and four dimensions, respectively. Therefore, the universality of
the coefficients $c_i$, as noted in sections \ref{sc:6DAnoms} and
\ref{sc:4DAnoms}, is essential to ensure this.

Let us close with some comments why the local Green--Schwarz mechanism
will always work within heterotic orbifold models; i.e.\ explain why
the local factorization of the orbifolds $T^6/\Intr_3$ or
$T^6/\Intr_4$ were no accidents. As discussed in this paper and in 
refs.\ \cite{Gmeiner:2002es,GrootNibbelink:2003gb}, the local shifts
at the orbifold fixed points should satisfy the appropriate modular
invariance requirements of string theory. However, for the zero mode
anomalies, it was demonstrated in ref.\ \cite{Schellekens:1987xh} that
modular invariance implies factorization. Since for pure orbifold
models, there is a direct identification between the zero mode anomaly
and the local anomalies at the fixed points, this implies that the
factorization holds at all fixed points separately. Naturally, this
local factorization continues to hold when Wilson lines are present,
since the factorization only depends on the modular invariance of the
local gauge shift at a given fixed point.

\section{Fayet--Iliopoulos tadpoles on $T^6/\Intr_4$}
\labl{sc:FItadp}

The appearance of anomalous $\U{1}$'s in a four dimensional
supersymmetric gauge theory is associated with Fayet--Iliopoulos
tadpoles for the auxiliary $D$ fields of the $\U{1}$ vector multiplets
being generated. For the heterotic string compactified on
$T^6/\Intr_3$ the local structure of such tadpoles were calculated in 
\cite{GrootNibbelink:2003gb}. Moreover, it was explicitly shown there
that these tadpoles for auxiliary field components are associated with
tadpoles for the internal gauge fields. Here we refrain from giving a
complete discussion of these different tadpoles on $T^6/\Intr_4$. We
simply give the expressions for the $D$ term tadpoles, as they can
straightforwardly be obtained from the general trace formulae
presented in section \ref{sc:traces}.

\begin{figure}
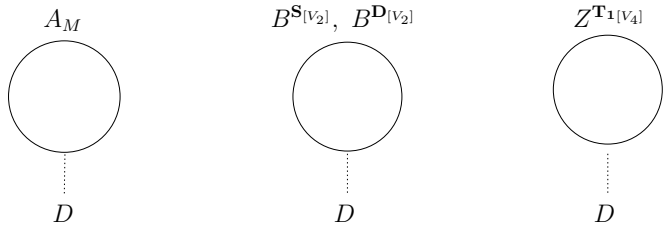

\[
\arry{ccc}{
\raisebox{0ex}{\scalebox{0.8}{\mbox{\input{FI_10D.pstex_t}}}}
\qquad & \qquad 
\raisebox{0ex}{\scalebox{0.8}{\mbox{\input{FI_6D.pstex_t}}}}
\qquad & \qquad 
\raisebox{0ex}{\scalebox{0.8}{\mbox{\input{FI_4D.pstex_t}}}}
}
\]
\caption{
Fayet--Iliopoulos tadpoles for $D$ generated by ten dimensional gauge
fields and six and four dimensional twisted states.  
} 
\labl{fig:FItadp}
\end{figure}

Using the four dimensional off--shell formulation of ten dimensional
super Yang--Mills \cite{Marcus:1983wb}, it is not difficult to see
that the possible diagrams that contribute to tadpoles are the ones
given in figure \ref{fig:FItadp}. In the loop they contain the ten
dimensional gauge fields and the six and four dimensional twisted
states. 

Since the only possible local anomalous $\U{1}$ generators are
$q_{v^4_{p\, q}}$ or $q_{v^4_{p\, q}}'$ defined in \eqref{defqq'}, we
only have to give tadpole expressions for those generators for fixed
point $\fZ^4_{p\, q}$: 
\equ{
L_{FI} =  D^I \int \frac{\d^4 p}{(2\pi)^4} 
\left\{ 
\frac {\frac {2}{16} \tr_{\rep{r}_{[v^4_{p\,q}]}}(H_I)}
{p^2 + \frac 14 \gD_1 + \frac 12 \gD_{23}} 
+ 
\frac {\frac {1}{4} \tr_{\rep{T_2}_{[v^4_{p\,q}]}}(H_I)}{p^2 + \frac 14 \gD_1} 
+ 
\frac {\tr_{\rep{T_1}_{[v^4_{p\,q}]}}(H_I)}{p^2} 
\right\} \gd^6( z - \fZ^4_{p\,q}). 
}
The factors $1/4$ and $1/2$ in front of the internal Laplacian  
$\gD_1 = \bder_1 \der_1$ and 
$\gD_{23} = \bder_2 \der_2 + \bder_3 \der_3$ 
are a consequence of the trace formulae, see \eqref{Tr10D} and
\eqref{Tr6D}.  We have written this expression in such a way, that the
relative contributions of the different terms at each of the fixed
points of $T^6/\Intr_4$, can directly be read off from table
\ref{tb:U1sums}.

If one uses a cut--off scheme to regularize the divergences here, one
finds quadratically divergent contributions for the anomalous
$\U{1}$'s. However, for all $\U{1}$ factors, anomalous or not,
at least logarithmically divergent tadpoles are generated. 
Due to supersymmetry similar tadpoles arise for the Cartan
directions of the internal gauge fields as well. Their
background will be similar to the one obtained in ref.\
\cite{GrootNibbelink:2003gb} for the $T^6/\Intr_3$ orbifold.

\section{Conclusions}
\labl{sc:concl}

In this paper we have investigated the local properties of
heterotic $\E{8}\times \E{8}'$ theory compactified on
$\Real^4\times T^6/\Intr_4$. Because this orbifold contains both, four
and six dimensional fixed hyper surfaces, the zero mode spectrum of the
theory can be rather complicated; certainly when Wilson lines are
present. However, at the fixed points locally
the structure of the heterotic string is very transparent: At the six
dimensional hyper surfaces all spectra are equivalent to one of two
possible $T^4/\Intr_2$ models, while at the four dimensional fixed
spaces there are essentially 12 different spectra possible.

We derived the local anomalies on the orbifolds $T^6/\Intr_4$,
$T^4/\Intr_2$ and  $T^2/\Intr_2$ by applying a general method to
compute traces over orbifold Hilbert 
spaces, developed in \cite{GrootNibbelink:2003gd}. The
calculation of the anomalies on $T^4/\Intr_2$
confirms the expectation that the anomaly structure of this orbifold
and of the six dimensional hyper surfaces of $T^6/\Intr_4$ are closely
related: However, an important numerical difference of a half was
found to be the result that one is a $\Intr_2$ and the other a
$\Intr_4$ orbifold. Some of the six dimensional hyper surfaces within
$\Real^4 \times T^6/\Intr_4$ are orbifolds themselves, namely
$T^2/\Intr_2$'s. The twisted states there give rise to anomalies in
four and six dimensions.

Collecting the anomaly contributions from the various sectors of
heterotic string theory on this orbifold, we found the following
results: The ten dimensional anomaly has a normalization factor of
$1/4$ w.r.t.\ the standard ten dimensional heterotic theory. 
The non--Abelian anomalies factorize at each of the six dimensional
fixed hyper surfaces separately. (Six dimensional Abelian anomalies
are absent for all models.) Similarly, the Abelian anomalies
factorize at the four dimensional fixed points, while non--Abelian
anomalies never arise there. These conclusions were obtained by
using the fact, that at each four or six dimensional fixed point only a
finite number of equivalent models can arise. Because of the universal
factorization involving the restriction of 
$\tr R^2 - \frac 1{30}\tr (iF)^2 - \frac 1{30}\tr (iF)^2$ to the
respective local gauge groups of the various fixed surfaces, the
Green--Schwarz mechanism is able to cancel all factorized anomalies, in
four, six and ten dimensions, locally at the same time. Using the
modular invariance of the local shifts, it can be understood that
factorization is implied \cite{Schellekens:1987xh}.

We computed the local Fayet--Iliopoulos tadpoles at the four dimensional fixed
points by using the general trace formulae of
\cite{GrootNibbelink:2003gd} again. For all local models involving
$\U{1}$ factors, such tadpoles 
are generated, even if those $\U{1}$ factors are not anomalous 
themselves. However, in that case the divergence of the
Fayet--Iliopoulos tadpole is only logarithmically instead of
quadratically.

\section*{Acknowlegdments}

SGN would like to thank the members of the department of theoretical 
physics of the Kyoto university for their kind hospitality. 
MH and MW would like to thank H.P.\ Nilles and S.\ F\"orste for valuable
discussions. 

SGN acknowledges the support of CITA and NSERC. The work of MH 
and MW has been supported in part by the European Community's Human
Potential Programme under contracts HPRN--CT--2000--00131 Quantum  
Spacetime, HPRN--CT--2000--00148 Physics Across the Present Energy
Frontier and HPRN--CT--2000--00152 Supersymmetry and the Early
Universe. TK is supported in part by the Grant-in-Aid for
Scientific Research from Ministry of Education, Science,
Sports and Culture of Japan (\#14540256).

\appendix
\def\theequation{\thesection.\arabic{equation}} 
\def\thetable{\thesection.\arabic{table}} 

\setcounter{equation}{0}
\setcounter{table}{0}
\section{Spinors in ten dimensions}
\labl{sc:spinlight}

This appendix provides some useful background for ten dimensional
spinors which are used in the main text of this work. 
(More details can be found in \cite{VanProeyen:1999ni,pol_2}.) 
We take the $(1,9)$ dimensional Clifford algebra generated by 
\equ{
\arry{l}{
\gG_0 = (i\gs_1,1,1,1,1),
\\[1ex]
\gG_1 = (\gs_2,1,1,1,1),
\\[1ex]
\gG_2 = (\gs_3,\gs_1,1,1,1),
\\[1ex]
\gG_3 = (\gs_3,\gs_2,1,1,1),
\\[1ex]
\gG_4 = (\gs_3,\gs_3,\gs_1,1,1),
}
\qquad 
\arry{l}{
\gG_5 = (\gs_3,\gs_3,\gs_2,1,1),
\\[1ex]
\gG_6 = (\gs_3,\gs_3,\gs_3,\gs_1,1),
\\[1ex]
\gG_7 = (\gs_3,\gs_3,\gs_3,\gs_2,1),
\\[1ex]
\gG_8 = (\gs_3,\gs_3,\gs_3,\gs_3,\gs_1),
\\[1ex]
\gG_9 = (\gs_3,\gs_3,\gs_3,\gs_3,\gs_2).
}
\labl{Cldecomp}
}
Here we have introduce a short hand notation for the tensor product of
five times the two dimensional Clifford space. These two dimensional
Clifford spaces are generated by the Pauli matrices $\gs_1$ and
$\gs_2$. The two dimensional chirality operator is $\gs_3$. An
explicit representation of these matrices reads
\equ{
\gs_1 = \pmtrx{0 & 1 \\ ~1 & 0},
\qquad 
\gs_2 = \pmtrx{0 & i \\ \mbox{-}i & 0},
\qquad 
\gs_3 = \pmtrx{1 & 0 \\ 0 & \mbox{-} 1}.
}
The matrices $\gs_1$ and $\gs_2$ can also be used as the charge
conjugation matrices $s_+ = \gs_1$ and 
$s_- = \gs_2 = \mbox{-}i \gs_3 s_+$  in two dimensions
\equ{
s_\pm\inv \, \gs_i \, s_\pm = \pm \gs_i^T, 
\ i = 1,2, 
\qquad 
s_\pm\inv\, \gs_3 \, s_\pm = - \gs_3^T,
\qquad
s_\pm\inv = s_\pm = s_\pm^\dag = - s_\pm^T. 
}
Notice that in two Euclidean dimensions one can only define Majorana
fermions  w.r.t.\ the charge conjugation $s_+$, because 
$(\get^{s_\pm})^{s_\pm} = \pm \get$, where $\get^{s_\pm} = s_\pm 
(\get^\dagger)^T$. Now, let $\get$ be a Majorana
fermion and denote two dimensional chiral spinors as $\get_\gk$, where
$\gs_3 \get_\gk = \gk \get_\gk$. The $s_+$ charge conjugates of these
spinors have opposite chirality: $(\get_\gk)^{s_+} = \get_{\mbox{-}\gk}$.

Using the basis of generators for the $(1,9)$ dimensional Clifford
algebra and the charge conjugations in two dimensions, it is not hard
to show that the charge conjugation matrices 
$C^\pm = (s_\pm, s_\mp, s_\pm, s_\mp, s_\pm)$ 
in ten dimensions have the properties
\equ{
(C^\pm)\inv \gG_M C^\pm = \pm \gG_M^T,
\quad 
(C^\pm)\inv \tgG C^\pm = - \tgG^T,
\quad 
(C^\pm)\inv = (C^\pm)^\dag 
= C^\pm = \pm (C^\pm)^T.
}
Here we have introduced the $(1,9)$ dimensional chirality operator 
$\tgG = (\gs_3, \gs_3, \gs_3, \gs_3, \gs_3)$. A basis for ten
dimensional spinors is given by  
\equ{
| S \rangle_{(1,9)} = \get_{2S_0} \otimes \ldots \otimes \get_{2S_4},  
\qquad 
\Id^i \otimes \gs_3 \otimes \Id^{4-i} \, | S \rangle_{(1,9)} = 
2 S_i \, | S \rangle_{(1,9)}, 
}
with $S = (S_i)$, $i = 0,\ldots 4$ and $S_i = \pm \frac 12$. The
Majorana conjugation of such a spinor reads 
\equ{
\Bigl( | S \rangle_{(1,9)}\Bigr)^{C_\pm} = 
\ga_\pm \,  | \mbox{-}S \rangle_{(1,9)},
\labl{BasisConj}
}
with $\ga_+ = - (2S_1)(2S_3)$ and $\ga_- = i (2S_0)(2S_2)(2S_4)$. 
However, a general spinor build out of this basis is not irreducible. 
In fact, from the properties above, it follows that in $(1,9)$
dimensions Majorana--Weyl fermions exist:
\equ{
\tgG \gch = \gb\, \gch, 
\qquad 
\gch^{C^\pm} = C^\pm \bgch^T = \gch, 
}
with chirality $\gb = \pm$. Since we encounter spinors of both
chiralities in the main text, we only solve the Majorana condition
explicitly: By going to light--cone gauge (with the assumption that the
spatial momentum vector is in the 1 direction) only the last four
spinor indices of $|S \rangle_{(1,9)}$ are relevant, and we may fix 
$S_0 = +1/2$, since by Majorana conjugation \eqref{BasisConj} we 
can always obtain $S_0 =-1/2$. This leads to the introduction of
a spinorial basis for $\SO{8}$ 
\equ{
| S \rangle = \get_{2S_1} \otimes \ldots \otimes \get_{2S_4}. 
\labl{SO8spin}
}
The relation with the $(1,9)$ dimensional spinors is therefore 
$|\pm\frac 12, \pm S \rangle_{(1,9)} = \get_\pm \otimes | \pm S \rangle$. 
Since we may fix $S_0 = 1/2$ by the Majorana condition, it follows
that a $(1,9)$ spinor of chirality $\pm$ is represented by $\SO{8}$
spinors $| S \rangle$ with $\prod (2S_i) = \pm$. (The definition of 
$| S \rangle$ can be naturally extended to any irreducible
representation of $\SO{8}$. For example, the vector 
representation is denoted by $| \underline{\pm 1, 0,0,0} \rangle$.)

\setcounter{equation}{0}
\setcounter{table}{0}
\section{Supersymmetric multiplets in six dimensions}
\labl{sc:6Dmulti}

This appendix is devoted to a brief exposition of
important supersymmetric multiplets in six dimensions. Following 
ref.\ \cite{Strathdee:1987jr} we classify the multiplets using the
little group $\SU{2}_+ \times \SU{2}_- = \SO{4} \subset \SO{1, 5}$ and
the $R$--symmetry group $\SU{2}_R$ of supersymmetry. (The subscript
$\pm$ refers to the $\Spin{4}$ chiralities.) On the light--cone
the following multiplets can be found: 
\begin{center}
\begin{tabular}{lclcl}
SUGRA & = & ({\bf 3}, {\bf 3}; {\bf 1}) + ({\bf 3}, {\bf 1}; {\bf 1}) 
&+& ({\bf 3}, {\bf 2}; {\bf 2}) \\
tensor & = & ({\bf 1}, {\bf 3}; {\bf 1}) + ({\bf 1}, {\bf 1}; {\bf 1})
&+& ({\bf 1}, {\bf 2}; {\bf 2}) \\
vector & = & ({\bf 2}, {\bf 2}; {\bf 1}) &+& ({\bf 2}, {\bf 1}; {\bf 2}) \\ 
half--hyper & = & ({\bf 1}, {\bf 1}; {\bf 2}) &+& ({\bf 1}, {\bf 2}; {\bf 1}) 
\end{tabular}
\end{center}
in terms of irreducible $\SU{2}_+ \times \SU{2}_- \times \SU{2}_R$
representations. Here the last terms of each row always contain the
fermionic super partners of the multiplet. The bosonic content
of the supergravity multiplet is the graviton $g_{mn}$ (with $m,n=2,\ldots 5$
spacetime indices in light--cone gauge) plus the
anti--self--dual part of the anti--symmetric tensor $B_{mn}$.  
The bosonic constituents of the tensor multiplet are the self--dual
part of the anti--symmetric tensor $B_{mn}$ and a dilaton $\gs$. 
From this we see that in supergravity models the supergravity
multiplet comes together with a tensor multiplet.

As some of these multiplets may come from ten dimensions upon
compactification, we briefly describe how these representations can be
described using the $\SO{8}$ weights discussed in section
\ref{sc:10Dlocal} and appendix \ref{sc:spinlight}. From the
branching rule 
\begin{equation}
\SO{8} \rightarrow \SO{4}_1 \times \SO{4}_2 \rightarrow
\SU{2}_+ \times \SU{2}_- \times \SU{2}_R \times \SU{2}_H
\labl{SO8branching}
\end{equation}
we get the following roots corresponding to $\SU{2}$'s: 
\(
\alpha_{+} = (1, 1, 0, 0), \; 
\alpha_{-} = (1, \mbox{-}1, 0, 0), \; 
\alpha_{R} = (0, 0, 1, 1), 
\) 
and $\alpha_{H} = (0, 0, 1,\mbox{-}1)$. The holonomy group $\SU{2}_H$ just
represents an internal symmetry from the six dimensional point of
view. (Since the holonomy $\SU{2}_H$ is broken to $\mathbb{Z}_2$ in
our orbifold limit, the representations under $\SU{2}_H$ give merely 
rise to phase factors and multiplicities.)

For a detailed description of the construction of supergravity,
tensor, vector and hyper multiplet systems in six dimensions, see
refs.\ \cite{Bergshoeff:1986mz,Kugo:2000hn} for example. Here we 
mainly focus on the fermionic properties of these multiplets as they
are important for the computation of six dimensional anomalies. 
To describe these properties we take the gamma matrices 
$\gG_m$, $m = 0, \ldots 5$, that generate the Clifford algebra 
in six dimensions. Using the basis given in \eqref{Cldecomp}, the
corresponding chirality and charge conjugation matrices can be represented by 
\equ{
\tgG_{6} = (\gs_3,\gs_3, \gs_3,1,1),
\qquad 
C_{6} = (s_-, s_+, s_-,1,1). 
}
Their properties can be summarized as 
\equ{
(C_{6})\inv \gG_m C_{6} = - \gG_m^T,
\quad 
(C_{6})\inv \tgG_{6} C_{6} = -\tgG_{6}^T, 
\quad 
(C_{6})\inv = (C_{6})^\dag 
= C_{6} = + (C_{6})^T.
}
We can define six dimensional chiral spinors as those having a
definite eigenvalue under $\tgG_6$. One cannot define 
Majorana fermions because for any spinor $\gz$
\equ{
(\gz^{C_6})^{C_6} = - \gz. 
\labl{doublConj}
}
However, we can define symplectic Majorana fermions  
\equ{
\gz^{C_6} = \gr\, \gz, 
\qquad 
(\gr^\dag)^T \gr = - \Id, 
\qquad
\gr^T = - \gr,
}
in terms of a matrix $\gr$. These two properties follow by 
demanding consistency with \eqref{doublConj} and the requirement that the 
kinetic Lagrangian for the fermions $\gz = (\gz^\ga)$ is a scalar
quantity. It follows that the number of indices $\ga$ is always
even. For the smallest choice of two, the matrix $\gr$ then takes the form
\equ{
\ge = -i \gs_2 = \pmtrx{ ~~0 & 1 \\ -1 & 0}. 
}
In general the matrix $\gr$ can always be brought to the form $\gr =
\Id \otimes \ge$. 
The essential properties of six dimensional spinors are therefore
encoded in their chirality and the corresponding matrix $\gr$. For the
relevant multiplets of this work, we have listed them in table
\ref{tb:6DFermions}.

\begin{table}
\begin{center}
\renewcommand{\arraystretch}{1.25}
\tabu{| l l | l | l |}{
\hline 
\multicolumn{4}{|c|}{
six dimensional supermultiplets}
\\ \hline \hline 
multiplet & content &  reality & chirality 
\\ \hline 
SUGRA  & $g_{mn}$, $B_{mn}$, $\gf$,   & 
$(\gps_m)^{C_6} = \ge\, \gps_m$ & $\tgG_6 \gps_m = \gps_m$ 
\\ 
+ tensor  & $\psi_m^i$, $\gl^i$ & $\gl^{C_6} = \ge\, \gl$ & 
$\tgG_6 \gl = - \gl$
\\ \hline 
vector & $A_m$, $\gch^i$ & $\gch^{C_6} = \ge\, \gch$ & 
$\tgG_6 \gch = \gch$
\\ \hline 
hyper & $q_i^\ga$, $\gz^\ga$ & $q^* = -\gr\, q \, \ge$ & - 
\\ 
& & $\gz^{C_6} = \gr\, \gz$ & $\tgG_6 \gz = - \gz$ 
\\ \hline 
}
\end{center}
\caption{
The most common six dimensional supergravity multiplets are given
together with the reality and chirality properties. The indices $i$
and $\ga$ are $\SU{2}_R$ and $\USp{2N}$ indices, respectively. 
}
\labl{tb:6DFermions}
\end{table}

The reality condition of the hyper multiplets appearing in table
\ref{tb:6DFermions} is of special interest in section
\ref{sc:6Dlocal}. If the fermions are in a real or complex
representation, then the number of components has to be doubled before
the reality condition can be imposed: Let $\gps$ be a chiral fermion
is such a representation, then the corresponding chiral
symplectic Majorana spinor is given by 
\equ{
\gz = \pmtrx{ \gps \\  - \gps^{C_6}}, 
\quad \text{with} \quad 
\gz^{C_6} = \pmtrx{ ~~0 & \Id \\ -\Id & 0} \gz.
} 
However, if the fermion $\gps$ is in a pseudo real representation,
then it can be directly used to form a special kind of hyper
multiplet: the so--called half--hyper multiplet. The reason for this
is that a pseudo real representation comes with a real
anti--symmetric matrix $\gr$ such that \cite{Georgi:1982jb,Slansky:1981yr}
\equ{
T^* = - \gr\, T \, \gr\inv, 
\qquad 
\gr^T = - \gr, 
\labl{pseudo}
}
for the representation matrix $T$ of the group.

\setcounter{equation}{0}
\setcounter{table}{0}
\section{$\boldsymbol{\E{8}}$ Weyl reflections and classification of
$\boldsymbol{\E{8}}$ shifts} 
\labl{sc:WeylReflect}

In the main text we have used extensively that the local shifts are
equivalent to only a limited set of standard shifts of both
$\Intr_2$ as well as $\Intr_4$ twists. Here we give a classification of the
possible shifts. The material presented in this appendix is related to
the $\Intr_3$ shift classification presented in 
\cite{Gmeiner:2002es,Casas:1989wu}. In section
\ref{sc:10Dlocal} we have given the $\E{8}$ roots as roots and weights
of the positive chiral spinor representation of $\SO{16}$. 
Since a gauge shift $v$ of a $\mathbb{Z}_N$ orbifold has to fulfill 
$N  v^I w_I  \equiv 0$ for all roots $w$, it follows that $N v$ is an
element of the $\E{8}$ root lattice, as this lattice is self--dual. 
Two $\E{8}$ gauge shifts $v$ and $v'$ are said to be equivalent, 
$v \simeq v'$, if  
\equ{
v' = v + u, ~ u \in \gG_8 
\quad \text{or} \quad
v' = W_{\ga}(v)  = v - (\ga, v) \ga.
}
where $W_\ga(v)$ denotes the Weyl reflection in root $\ga$ of
$\E{8}$. Since all $\E{8}$ roots have length 2, it follows that for a
Weyl reflection ${v'}^2 = v^2$. 

A useful application are the Weyl reflections at the
$SO(16)$ roots
\equ{
\label{signsaway}
(~v_1, ~v_2, ~v_3, \ldots) \simeq 
W_{(~1, \pm 1, ~0^6)}(~v_1, ~v_2, ~v_3, \ldots) = 
(\mp v_2, \mp v_1, ~v_3, \ldots).
}
Hence we see that by interchanging two shift elements, or replacing two
shift elements by minus those elements equivalent shifts are
obtained. In particular, if a shift has at least one zero, the sign of
all other entries is irrelevant.

\subsection{Classification of $\boldsymbol{\Intr_2}$ gauge shifts}
\labl{sc:Z2class}

First of all, if $V_2$ is an element of the $\E{8}$ root lattice, the
gauge group will not be broken, and hence be equal to $\E{8}$. Since
the $\E{8}$ lattice is even, such a lattice vector fulfills $2 (V_2)^2
= 0 \mod 4$.
Assume that $V_2$ is not an $\E{8}$ root lattice vector. 
To classify these $\Intr_2$ gauge shifts we note that for all $\E{8}$
roots $\ga$ we have $(V_2 + \ga)^2 = V_2^2 \mod 1$, since 
$(\ga, V_2) = 0, 1/2 \mod 1$ and $\ga^2 = 2$. From this it follows
that for equivalent $\Intr_2$ gauge shifts $V_2^2 \mod 1$ is equal,
since the sum of squared entries is always invariant under Weyl
reflections. This completes the description of the $\Intr_2$ gauge
shift classification, the three possibilities are given in 
table \ref{tb:class-2-U}. A standard shift can be defined by a shift
with maximal number of zeros and all entries positive, which is the
form we used in that table.

\subsection{Classification of $\boldsymbol{\Intr_4}$ gauge shifts}
\labl{sc:Z4class}

For the classification of $\Intr_4$ gauge shifts, we first need to
bring them to a standard form: 
The entries of $4v$ can either all be half-integer, or all be
integer. Since all inequivalent models can be computed using only
the latter type of gauge shifts, we will not consider half-integer
entries here. 
Since $(2, 0^7)$ and its permutations are the sums of two roots of
$\SO{16}$, we infer that the integer valued entries of $4v$ can be
restricted to $4v^I =  -3, \ldots, 4$. In fact, we may even assume
that either $4v = (4, 0^7)$ or that no entry $4v^I$ is equal to $4$:
If there are two or more entries equal to $4$, then by adding the
$\SO{16}$ root with $-1$ at two of these entries, they become zero.  
If there is just one entry equal to $4$, then either $4v = (4, 0^7)$ 
or there is another entry of $4v^I = -3, \ldots, 3$, for which, by
adding a $\SO{16}$ root again, we can make the $4$ entry $0$, and map
$4v^I$ back into $4v^I = -3, \ldots, 3$. (We have assumed, that if the shift
contains more than one non--vanishing entries, this procedure
has been applied throughout the paper to set all entries 
$4v^I \in \{-3, -2, -1, 0, 1, 2, 3\}$.)

From now on, we would like to exploit the following conditions: 1)
Since $4v$ is in the $\E{8}$ root lattice, 
the number of odd entries of $4v$ is always even. 
2) If there is at least one entry $4v^J = 0$ it follows from
(\ref{signsaway}), that signs do not matter anymore; we take them
positive. 3) A pair of entries $\pm 3$ can always be mapped to a pair
$\pm 1$ by addition of an $\SO{16}$ root, so that we can restrict to
at most one entry of $3$. 4) If there is a entry of $2$ in a gauge shift
$4v$ with at least one zero, by adding
a $\SO{16}$ root with $-1$ at this entry, the $2$ gets mapped to $-2$
which is with the help of 2) equivalent to $2$ again, so that all
other non--vanishing entries of $4v$ are equivalent modulo $4$. 
The standard form of a gauge shift is defined to be that form with the
least possible numbers of $3$'s and $2$'s in $4v$. And we require that all
entries are positive. This is always possible except when all entries
of $4v$ are $\pm 1$. If the number of minus signs is even, they can be made
all positive, while in the odd case, one has to keep one $-1$. 
The number of $3$'s and $2$'s can often be reduced by subtracting a
spinorial root, for example we have $(3,1^7)/4 \simeq (1^7,-1)/4$.

Once in this standard form, we can again consider the sum of the
square of the entries of the shift vector. This classifies the shifts
almost uniquely, expect when $(4v)^2 = 8$. In this case there
are two possibilities, but the sum of the entries of $2v \mod 2$ can be used
to distinguish between both possibilities. The results are summarized
in table  \ref{tb:class-4-U}.

\bibliographystyle{paper.bst}
{\small
\bibliography{paper}

\providecommand{\href}[2]{#2}\begingroup\raggedright\begin{thebibliography}{10}

\bibitem{dixon_85}
L.~Dixon, J.~A. Harvey, C.~Vafa, and E.~Witten ``Strings on orbifolds'' {\em
  Nucl. Phys.} {\bf B261} (1985)
678--686.

\bibitem{Dixon:1986jc}
L.~J. Dixon, J.~A. Harvey, C.~Vafa, and E.~Witten ``Strings on orbifolds. 2''
  {\em Nucl. Phys.} {\bf B274} (1986)
285--314.

\bibitem{ibanez_87}
L.~E. Ibanez, H.~P. Nilles, and F.~Quevedo ``Orbifolds and {W}ilson lines''
  {\em Phys. Lett.} {\bf B187} (1987)
25--32.

\bibitem{Ibanez:1988pj}
L.~E. Ibanez, J.~Mas, H.-P. Nilles, and F.~Quevedo ``Heterotic strings in
  symmetric and asymmetric orbifold backgrounds'' {\em Nucl. Phys.} {\bf B301}
  (1988)
157.

\bibitem{Font:1988tp}
A.~Font, L.~E. Ibanez, H.~P. Nilles, and F.~Quevedo ``Degenerate orbifolds''
  {\em Nucl. Phys.} {\bf B307} (1988)
109.

\bibitem{Arkani-Hamed:2001is}
N.~Arkani-Hamed, A.~G. Cohen, and H.~Georgi ``Anomalies on orbifolds'' {\em
  Phys. Lett.} {\bf B516} (2001) 395--402
\href{http://arXiv.org/abs/hep-th/0103135}{{\tt hep-th/0103135}}.

\bibitem{Scrucca:2001eb}
C.~A. Scrucca, M.~Serone, L.~Silvestrini, and F.~Zwirner ``Anomalies in
  orbifold field theories'' {\em Phys. Lett.} {\bf B525} (2002) 169--174
\href{http://arXiv.org/abs/hep-th/0110073}{{\tt hep-th/0110073}}.

\bibitem{Barbieri:2002ic}
R.~Barbieri, R.~Contino, P.~Creminelli, R.~Rattazzi, and C.~A. Scrucca
  ``Anomalies, {F}ayet-{I}liopoulos terms and the consistency of orbifold field
  theories'' {\em Phys. Rev.} {\bf D66} (2002) 024025
\href{http://www.arXiv.org/abs/hep-th/0203039}{{\tt hep-th/0203039}}.

\bibitem{Pilo:2002hu}
L.~Pilo and A.~Riotto ``On anomalies in orbifold theories''
\href{http://arXiv.org/abs/hep-th/0202144}{{\tt hep-th/0202144}}.

\bibitem{GrootNibbelink:2002qp}
S.~Groot~Nibbelink, H.~P. Nilles, and M.~Olechowski ``Instabilities of bulk
  fields and anomalies on orbifolds'' {\em Nucl. Phys.} {\bf B640} (2002)
  171--201
\href{http://www.arXiv.org/abs/hep-th/0205012}{{\tt hep-th/0205012}}.

\bibitem{Asaka:2002my}
T.~Asaka, W.~Buchmuller, and L.~Covi ``Bulk and brane anomalies in six
  dimensions'' {\em Nucl. Phys.} {\bf B648} (2003) 231--253
\href{http://www.arXiv.org/abs/hep-ph/0209144}{{\tt hep-ph/0209144}}.

\bibitem{vonGersdorff:2003dt}
G.~von Gersdorff and M.~Quiros ``Localized anomalies in orbifold gauge
  theories''
\href{http://www.arXiv.org/abs/hep-th/0305024}{{\tt hep-th/0305024}}.

\bibitem{GrootNibbelink:2003gd}
S.~Groot~Nibbelink ``Traces on orbifolds: {A}nomalies and one-loop amplitudes''
  {\em JHEP} {\bf 07} (2003) 011
\href{http://www.arXiv.org/abs/hep-th/0305139}{{\tt hep-th/0305139}}.

\bibitem{Gmeiner:2002es}
F.~Gmeiner, S.~Groot~Nibbelink, H.~P. Nilles, M.~Olechowski, and M.~Walter
  ``Localized anomalies in heterotic orbifolds'' {\em Nucl. Phys.} {\bf B648}
  (2003) 35--68
\href{http://www.arXiv.org/abs/hep-th/0208146}{{\tt hep-th/0208146}}.

\bibitem{GrootNibbelink:2003gb}
S.~Groot~Nibbelink, H.~P. Nilles, M.~Olechowski, and M.~G.~A. Walter
  ``Localized tadpoles of anomalous heterotic {U(1)}'s''
\href{http://www.arXiv.org/abs/hep-th/0303101}{{\tt hep-th/0303101}}.

\bibitem{Green:1984sg}
M.~B. Green and J.~H. Schwarz ``Anomaly cancellation in supersymmetric {D=10}
  gauge theory and superstring theory'' {\em Phys. Lett.} {\bf B149} (1984)
117--122.

\bibitem{Dine:1987xk}
M.~Dine, N.~Seiberg, and E.~Witten ``{F}ayet-{I}liopoulos terms in string
  theory'' {\em Nucl. Phys.} {\bf B289} (1987)
589.

\bibitem{Atick:1987gy}
J.~J. Atick, L.~J. Dixon, and A.~Sen ``String calculation of
  {F}ayet-{I}liopoulos {D} terms in arbitrary supersymmetric
  compactifications'' {\em Nucl. Phys.} {\bf B292} (1987)
109--149.

\bibitem{Dine:1987gj}
M.~Dine, I.~Ichinose, and N.~Seiberg ``{F} terms and {D} terms in string
  theory'' {\em Nucl. Phys.} {\bf B293} (1987)
253.

\bibitem{Ghilencea:2001bw}
D.~M. Ghilencea, S.~Groot~Nibbelink, and H.~P. Nilles ``Gauge corrections and
  {FI}-term in {5D KK} theories'' {\em Nucl. Phys.} {\bf B619} (2001) 385--395
\href{http://arXiv.org/abs/hep-th/0108184}{{\tt hep-th/0108184}}.

\bibitem{Barbieri:2001cz}
R.~Barbieri, L.~J. Hall, and Y.~Nomura ``A constrained standard model: Effects
  of {F}ayet-{I}liopoulos terms''
\href{http://www.arXiv.org/abs/hep-ph/0110102}{{\tt hep-ph/0110102}}.

\bibitem{GrootNibbelink:2002wv}
S.~Groot~Nibbelink, H.~P. Nilles, and M.~Olechowski ``Spontaneous localization
  of bulk matter fields'' {\em Phys. Lett.} {\bf B536} (2002) 270--276
\href{http://www.arXiv.org/abs/hep-th/0203055}{{\tt hep-th/0203055}}.

\bibitem{Ruan:2000uf}
Y.~Ruan ``Discrete torsion and twisted orbifold cohomology''
\href{http://www.arXiv.org/abs/math.ag/0005299}{{\tt math.ag/0005299}}.

\bibitem{Kobayashi:1991mi}
T.~Kobayashi and N.~Ohtsubo ``Analysis on the {W}ilson lines of {Z(N)} orbifold
  models'' {\em Phys. Lett.} {\bf B257} (1991)
56--62.

\bibitem{Kobayashi:1994rp}
T.~Kobayashi and N.~Ohtsubo ``Geometrical aspects of {Z(N)} orbifold
  phenomenology'' {\em Int. J. Mod. Phys.} {\bf A9} (1994)
87--126.

\bibitem{pol_2}
J.~Polchinski {\em String theory vol. 2: Superstring theory and beyond}.
\newblock Cambridge, Uk: Univ. Pr. 531 P. (Cambridge Monographs On Mathematical
  Physics) 1998.

\bibitem{DiFrancesco:1997nk}
P.~Di~Francesco, P.~Mathieu, and D.~Senechal ``Conformal field theory''. New
  York, USA: Springer (1997) 890 p.

\bibitem{Katsuki:1989qz}
Y.~Katsuki, Y.~Kawamura, T.~Kobayashi, N.~Ohtsubo, Y.~Ono, and K.~Tanioka
  ``{$Z(4)$} and {$Z(6)$} orbifold models'' {\em Phys. Lett.} {\bf B218} (1989)
  169.

\bibitem{Kawamura:1996zu}
Y.~Kawamura and T.~Kobayashi ``Flat directions in {$Z(2N)$} orbifold models''
  {\em Nucl. Phys.} {\bf B481} (1996) 539--576
\href{http://www.arXiv.org/abs/hep-th/9606189}{{\tt hep-th/9606189}}.

\bibitem{Fujikawa:1979ay}
K.~Fujikawa ``Path integral measure for gauge invariant fermion theories'' {\em
  Phys. Rev. Lett.} {\bf 42} (1979)
1195.

\bibitem{Fujikawa:1980eg}
K.~Fujikawa ``Path integral for gauge theories with fermions'' {\em Phys. Rev.}
  {\bf D21} (1980)
2848.

\bibitem{Zumino:1984rz}
B.~Zumino, Y.-S. Wu, and A.~Zee ``Chiral anomalies, higher dimensions, and
  differential geometry'' {\em Nucl. Phys.} {\bf B239} (1984)
477--507.

\bibitem{Bardeen:1984pm}
W.~A. Bardeen and B.~Zumino ``Consistent and covariant anomalies in gauge and
  gravitational theories'' {\em Nucl. Phys.} {\bf B244} (1984)
421.

\bibitem{Alvarez-Gaume:1984ig}
L.~Alvarez-Gaume and E.~Witten ``Gravitational anomalies'' {\em Nucl. Phys.}
  {\bf B234} (1984)
269.

\bibitem{Nakahara:1990th}
M.~Nakahara ``Geometry, topology and physics''. Bristol, UK: Hilger (1990) 505
  p. (Graduate student series in physics).

\bibitem{gsw_2}
M.~B. Green, J.~H. Schwarz, and E.~Witten {\em Superstring theory vol. 2:
  {L}oop amplitudes, anomalies and phenomenology}.
\newblock Cambridge, Uk: Univ. Pr. 596 P. (Cambridge Monographs On Mathematical
  Physics) 1987.

\bibitem{Green:1985bx}
M.~B. Green, J.~H. Schwarz, and P.~C. West ``Anomaly free chiral theories in
  six-dimensions'' {\em Nucl. Phys.} {\bf B254} (1985)
327--348.

\bibitem{Erler:1994zy}
J.~Erler ``Anomaly cancellation in six-dimensions'' {\em J. Math. Phys.} {\bf
  35} (1994) 1819--1833
\href{http://www.arXiv.org/abs/hep-th/9304104}{{\tt hep-th/9304104}}.

\bibitem{vanRitbergen:1998pn}
T.~van Ritbergen, A.~N. Schellekens, and J.~A.~M. Vermaseren ``Group theory
  factors for {F}eynman diagrams'' {\em Int. J. Mod. Phys.} {\bf A14} (1999)
  41--96
\href{http://www.arXiv.org/abs/hep-ph/9802376}{{\tt hep-ph/9802376}}.

\bibitem{Schellekens:1987xh}
A.~N. Schellekens and N.~P. Warner ``Anomalies, characters and strings'' {\em
  Nucl. Phys.} {\bf B287} (1987)
317.

\bibitem{Kobayashi:1997pb}
T.~Kobayashi and H.~Nakano ``Anomalous {U(1)} symmetry in orbifold string
  models'' {\em Nucl. Phys.} {\bf B496} (1997) 103--131
\href{http://arXiv.org/abs/hep-th/9612066}{{\tt hep-th/9612066}}.

\bibitem{Chamseddine:1981ez}
A.~H. Chamseddine ``Interacting supergravity in ten-dimensions: The role of the
  six - index gauge field'' {\em Phys. Rev.} {\bf D24} (1981)
3065.

\bibitem{Bergshoeff:1982um}
E.~Bergshoeff, M.~de~Roo, B.~de~Wit, and P.~van Nieuwenhuizen ``Ten-dimensional
  {M}axwell-{E}instein supergravity, its currents, and the issue of its
  auxiliary fields'' {\em Nucl. Phys.} {\bf B195} (1982)
97--136.

\bibitem{Chapline:1983ww}
G.~F. Chapline and N.~S. Manton ``Unification of {Y}ang-{M}ills theory and
  supergravity in ten dimensions'' {\em Phys. Lett.} {\bf B120} (1983)
105--109.

\bibitem{Marcus:1983wb}
N.~Marcus, A.~Sagnotti, and W.~Siegel ``Ten-dimensional supersymmetric
  {Y}ang-{M}ills theory in terms of four-dimensional superfields'' {\em Nucl.
  Phys.} {\bf B224} (1983)
159.

\bibitem{VanProeyen:1999ni}
A.~Van~Proeyen ``Tools for supersymmetry''
\href{http://arXiv.org/abs/hep-th/9910030}{{\tt hep-th/9910030}}.

\bibitem{Strathdee:1987jr}
J.~Strathdee ``Extended {P}oincare supersymmetry'' {\em Int. J. Mod. Phys.}
  {\bf A2} (1987)
273.

\bibitem{Bergshoeff:1986mz}
E.~Bergshoeff, E.~Sezgin, and A.~Van~Proeyen ``Superconformal tensor calculus
  and matter couplings in six- dimensions'' {\em Nucl. Phys.} {\bf B264} (1986)
653.

\bibitem{Kugo:2000hn}
T.~Kugo and K.~Ohashi ``Supergravity tensor calculus in {5D} from {6D}'' {\em
  Prog. Theor. Phys.} {\bf 104} (2000) 835--865
\href{http://www.arXiv.org/abs/hep-ph/0006231}{{\tt hep-ph/0006231}}.

\bibitem{Georgi:1982jb}
H.~Georgi ``Lie algebras in particle physics. {F}rom isospin to unified
  theories'' {\em Front. Phys.} {\bf 54} (1982)
1--255.

\bibitem{Slansky:1981yr}
R.~Slansky ``Group theory for unified model building'' {\em Phys. Rept.} {\bf
  79} (1981)
1--128.

\bibitem{Casas:1989wu}
J.~A. Casas, M.~Mondragon, and C.~Munoz ``Reducing the number of candidates to
  standard model in the {Z(3)} orbifold'' {\em Phys. Lett.} {\bf B230} (1989)
63.

\end{thebibliography}\endgroup
}

\end{document}